\documentstyle[pra,aps]{revtex}

\begin{document}
\title{Cavity induced modifications to the resonance fluorescence and probe
absorption of a laser-dressed V atom}
\author{Jin-Sheng Peng$^{1,2}$, Gao-Xiang Li$^1$, Peng Zhou$^2$, and S. Swain$^2$}
\address{$^1$Department of Physics, \\
Huazhong Normal University, Wuhan 430079, China,\\
$^2$Department of Applied Mathematics and Theoretical Physics, \\
The Queen's University of Belfast, Belfast BT7 1NN, UK}
\date{}
\maketitle

\begin{abstract}
A cavity-modified master equation is derived for a coherently driven, V-type
three-level atom coupled to a single-mode cavity in the bad cavity limit. We
show that population inversion in both the bare and dressed-state bases may
be achieved, originating from the enhancement of the atom-cavity interaction
when the cavity is resonant with an atomic dressed-state transition. The
atomic populations in the dressed state representation are analysed in terms
of the cavity-modified transition rates. The atomic fluorescence spectrum
and probe absorption spectrum also investigated, and it is found that the
spectral profiles may be controlled by adjusting the cavity frequency. Peak
suppression and line narrowing occur under appropriate conditions.
\end{abstract}


\section{Introduction}

A major interest of modern quantum optics is to devise ways to modify and
control the radiative properties of atoms. This may be achieved by changing
the environment, so that the atoms interact with a modified set of vacuum
modes. One such modified vacuum is provided by the cavity environment \cite
{s1}, where the electromagnetic modes are concentrated around the cavity
resonant frequency. The coupling of the atoms to the modified
electromagnetic vacuum is therefore frequency dependent. For an excited atom
located inside such a cavity, the cavity mode is the only one available to
the atom for emission. If the atomic transition is in resonance with the
cavity, the spontaneous emission rate into the particular cavity mode is
enhanced \cite{s2}; otherwise, it is inhibited \cite{s3}. When the atom is
strongly driven by a laser field, the atom-laser system may be considered to
form a new dressed atom \cite{s4,s5} whose energy-level structure is
intensity dependent. For such a coherently driven two-level atom placed
inside a cavity, theoretical investigations have predicted a
phenomenological richness which is not found in the absence of the strong
driving field---for example, dynamical suppression of the spontaneous
emission rate \cite{s6,s7}, population inversion in both bare and dressed
states basis \cite{s7,s8}, and distortion and narrowing of the Mollow
triplet \cite{s7,s9}. All these features are very sensitive to the cavity
resonance frequency because of the cavity enhancement of the dressed atomic
transitions.

Recently, Lange and Walther \cite{s10} have observed the dynamical
suppression of spontaneous emission in a microwave cavity. In the
optical-frequency regime, Zhu {\em et al.} \cite{s11} have also reported
experimental studies of the effects of cavity detuning on the radiative
properties of a coherently driven two-level atom. They have shown that the
atomic fluorescence of a strongly driven two-level atom is enhanced when the
cavity frequency is tuned to one of the sidebands of the Mollow fluorescence
triplet, whereas it is inhibited by tuning to the other sideband. The
enhancement of atomic resonance fluorescence at one sideband is a direct
demonstration of population inversion.

In this paper, we investigate the dynamical modification of the resonance
fluorescence of a coherently driven V-type three-level atom coupled to a
frequency-tunable, single-mode cavity in the bad cavity limit. We
demonstrate that the atomic populations, fluorescence spectrum and
absorption spectrum can be strongly controlled and manipulated by tuning the
cavity frequency. In Section 2, we derive a cavity-modified master equation
for the atomic density-matrix operator from the full master equation by
adiabatically eliminating the cavity variables in the bad cavity limit. For
simplicity, we restrict attention to the situation where the laser frequency
is tuned to the mean Bohr frequency of the excited states. The results
obtained here are basic to the whole paper, and the equations derived in
this section are used to calculate {\em all} the numerical results
presented. The first part of Section 3 is devoted to discussing the atomic
population distribution in the bare state representation. It is pointed out
that significant population inversions can be achieved. In the second part
of this section, we analyse the equations for the populations in the dressed
state basis, with particular emphasis on the high field limit. We find that
the dressed state populations obey rate equations, with atomic transition
rates that are strongly dependent on the Rabi frequency and the cavity
resonance frequency. Under certain conditions, the population may also be
inverted in the dressed-state basis. The plots of the dressed state
populations against the cavity frequency--laser frequency detuning are used
to provide a semi-quantitative understanding of the phenomena described in
the subsequent sections. In Section 4 we study the cavity effects on the
resonance fluorescence spectra of this system. It is shown that dynamical
control of the populations and fluorescence spectrum by adjustment of the
cavity resonance frequency and Rabi frequency is possible. Peak suppression
and line narrowing phenomena are also revealed. In Section 5 we briefly
consider the absorption of a weak tunable probe field transmitted through
this system, and in the last section we give a summary.

\section{The cavity modified master equation}

Consider a V-configuration atom consisting of two excited states $|1\rangle $
and $|2\rangle $ coupled to a ground state $|0\rangle $ by a single-mode
cavity field of frequency $\omega _{C}$ and a laser field with frequency $%
\omega _{L}$, as shown in Fig. 1. The cavity mode is described by the
annihilation and creation operators $a$ and $a^{\dag }$, while the atom is
represented by the operators $A_{lk}\equiv |l\rangle \langle
k|\;(l,\,k=0,\,1,\,2)$. In the frame rotating at the frequency $\omega _{L},$
and within the rotating wave approximation, the master equation for the
density matrix operator $\rho $ of the combined atom-cavity system is
\begin{equation}
\dot{\rho}_{T}=-i[H_{A}+H_{C}+H_{I},\rho _{T}]+{\cal L}_{A}\rho _{T}+{\cal L}%
_{C}\rho _{T}  \label{master1}
\end{equation}
where
\begin{mathletters}
\begin{eqnarray}
&&H_{A}=(\Delta -\omega _{21})A_{11}+\Delta A_{22}+\Omega
_{2}(A_{02}+A_{20})+\Omega _{1}(A_{01}+A_{10}), \\
&&H_{C}=\delta a^{\dag }a, \\
&&H_{I}=g_{2}(a^{\dag }A_{02}+A_{20}a)+g_{1}(a^{\dag }A_{01}+A_{10}a), \\
&&{\cal L}_{A}\rho _{T}=\frac{\gamma _{1}}{2}(2A_{01}\rho _{T}A_{10}-\rho
_{T}A_{11}-A_{11}\rho _{T})+\frac{\gamma _{2}}{2}(2A_{02}\rho
_{T}A_{20}-\rho _{T}A_{22}-A_{22}\rho _{T}), \\
&&{\cal L}_{C}\rho _{T}=\kappa (2a\rho _{T}a^{\dag }-a^{\dag }a\rho
_{T}-\rho _{T}a^{\dag }a),
\end{eqnarray}
with
\end{mathletters}
\begin{equation}
\omega _{21}=\omega _{2}-\omega _{1},\;\;\;\delta =\omega _{C}-\omega
_{L}\;\;\;\text{and}\;\;\;\Delta =\omega _{2}-\omega _{L}.
\end{equation}
Here $H_{A}$ and $H_{C}$ describe the coherently driven atom and the cavity
respectively, and $H_{I}$ represents the interaction between the atom and
the cavity mode. The Rabi frequency $\Omega _{j}$ relates to the atomic
transitions $|j\rangle \leftrightarrow |0\rangle $ $\left( j=1,2\right) $
under the action of the driving laser field, and $g_{j}$ is the coupling
constant between the atom and the cavity mode associated with the same
transition. ${\cal L}_{C}\rho _{T}$ and ${\cal L}_{A}\rho _{T}$ describe
respectively the damping of the cavity field by a standard vacuum reservoir,
and the atomic damping to background modes other than the privileged cavity
modes. $\gamma _{1}$ and $\gamma _{2}$ are just the spontaneous decay
constants of the levels $|1\rangle $ and $|2\rangle $. Here we also assume
that the atomic dipole moments ${\bf d}_{10}$ and ${\bf d}_{20}$ are
orthogonal to each other, so that there is no spontaneously generated
quantum interference \cite{s12} resulting from the cross coupling between
the transitions $|1\rangle \leftrightarrow |0\rangle $ and $|2\rangle
\leftrightarrow |0\rangle $. For simplicity in the resulting expressions, we
assume $\Omega _{1}=\Omega _{2}=\Omega $, $g_{1}=g_{2}=g$, $\gamma
_{1}=\gamma _{2}=\gamma $, and $\Delta =\omega _{21}/2$ in what follows.

We assume that the atom-cavity coupling is weak and the cavity has a low-Q
value, so that
\begin{equation}
\kappa \gg g\gg \gamma  \label{bc}
\end{equation}
(the bad cavity limit). This condition implies that the cavity mode response
to the standard vacuum reservoir is much faster than that produced by its
interaction with the atom. Then the atom always experiences the cavity mode
in the state induced by the vacuum reservoir, and this permits one to
eliminate the variables containing the cavity field operators adiabatically,
giving rise to a reduced master equation for the atomic variables only. As
the derivation is tedious, we refer readers to \cite{s7,s10}, and here only
outline the key points.

We temporarily disregard ${\cal L}_{A}\rho _{T}$ in the elimination of the
cavity-mode, since it unchanged by these operations. First we perform a
canonical transformation to the atom-cavity interaction picture (\ref
{master1}) by

\begin{equation}
\tilde{\rho}_T=e^{i(H_{A}+H_{C})t}\rho_T e^{-i(H_{A}+H_{C})t}.
\end{equation}
The master equation then takes the form

\begin{equation}
\partial _{t}\left( e^{-{\cal L}_{C}t}\tilde{\rho}_{T}\right) =-ie^{-{\cal L}%
_{C}t}[\tilde{H}_{I}(t),\tilde{\rho}_{T}],  \label{tramaster}
\end{equation}
where $\tilde{H}_{I}(t)=g[\tilde{D}(t)a^{\dag }\exp (i\delta t)+h.c.]$, with
$\tilde{D}(t)=\exp (iH_{A}t)D\exp (-iH_{A}t)$ and $D=A_{01}+A_{02}$. We next
introduce the operator $\chi $

\begin{equation}
\chi =e^{-{\cal L}_{C}t}\tilde{\rho}_{T},
\end{equation}
which, according to Eq. (\ref{tramaster}), obeys the equation

\begin{eqnarray}
\dot{\chi}(t) &=&-ige^{\kappa t}\left\{\left[a^{\dag },\,\tilde{D}(t)\chi (t)%
\right]e^{i\delta t}+\left[a,\,\chi (t)\tilde{D}^{\dag}(t)\right]e^{-i\delta
t}\right\}  \nonumber \\
&&-ige^{-\kappa t}\left\{\left[\tilde{D}(t),\,\chi (t)a^{\dag }\right]%
e^{i\delta t}+ \left[\tilde{D}^{\dag}(t),\,a\chi (t)\right]e^{-i\delta
t}\right\}.  \label{JC-like}
\end{eqnarray}

Due to the smallness of the coupling constant $g$, we can perform a
second-order perturbation calculation with respect to $g$ by means of
standard projection operator techniques. Noting that
\begin{equation}
\mbox{Tr}_{C}\chi (t)\equiv \mbox{Tr}_{C}\tilde{\rho}_T(t)\equiv \tilde{\rho}%
(t),
\end{equation}
we can trace out the cavity variables to obtain the master equation for the
reduced density matrix operator $\tilde{\rho}$ of the atom. Under the
Born-Markovian approximation, the resulting master equation is of the form

\[
\tilde{\rho}(t)=-g^{2}\int_{0}^{\infty }\left\{ e^{-(\kappa +i\delta )\tau }[%
\tilde{D}^{\dag }(t)\tilde{D}(t-\tau )\tilde{\rho}(t)-\tilde{D}(t-\tau )%
\tilde{\rho}(t)\tilde{D}^{\dag }(t)]+h.c.\right\} d\tau .
\]
Finally transforming $\tilde{\rho}$ back to the original picture via $\rho
=\exp (-iH_{A}t)\tilde{\rho}\exp (iH_{A}t)$, and restoring the ${\cal L}%
_{A}\rho $ contribution, we express the reduced master equation for the
atomic variables as

\begin{eqnarray}
\dot{\rho} &=&-i\left[ H_{A},\;\rho \right]  \nonumber \\
&&+\frac{\gamma _{c}}{2}\left( D\rho S^{\dag }+S\rho D^{\dag }-D^{\dag
}S\rho -\rho S^{\dag }D\right)  \nonumber \\
&&+\frac{\gamma }{2}\left( 2A_{01}\rho A_{10}-\rho A_{11}-A_{11}\rho \right)
+\frac{\gamma }{2}\left( 2A_{02}\rho A_{20}-\rho A_{22}-A_{22}\rho \right)
\label{mast}
\end{eqnarray}
where $\gamma _{c}=2g^{2}/\kappa $ specifies the emission rate of the atom
into the cavity mode, and
\begin{eqnarray}
S &=&\kappa \int_{0}^{\infty }e^{-(\kappa +i\delta )\tau }\tilde{D}(-\tau
)d\tau  \nonumber \\
&=&\beta _{0}A_{00}+\beta _{1}A_{11}+\beta _{2}A_{22}+\beta _{3}A_{10}+\beta
_{4}A_{01}+\beta _{5}A_{20}+\beta _{6}A_{02}+\beta _{7}A_{21}+\beta
_{8}A_{12},
\end{eqnarray}

where the coefficients $\beta _{i}$ ($i=0,1,....8$) are given by

\begin{equation}
\left[
\begin{array}{l}
\beta _{0} \\
\beta _{1} \\
\beta _{2} \\
\beta _{3} \\
\beta _{4} \\
\beta _{5} \\
\beta _{6} \\
\beta _{7} \\
\beta _{8}
\end{array}
\right] =\left[
\begin{array}{ccccc}
0 & 2\eta \varepsilon ^{2} & -2\eta \varepsilon ^{2} & 8\eta ^{3} & -8\eta
^{3} \\
-2\eta \varepsilon & \eta \varepsilon \left( 1-\varepsilon \right) & \eta
\varepsilon \left( 1+\varepsilon \right) & -\eta \left( 1-\varepsilon
^{2}\right) /2 & \eta \left( 1-\varepsilon ^{2}\right) /2 \\
2\eta \varepsilon & -\eta \varepsilon \left( 1+\varepsilon \right) & -\eta
\varepsilon \left( 1-\varepsilon \right) & -\eta \left( 1-\varepsilon
^{2}\right) /2 & \eta \left( 1-\varepsilon ^{2}\right) /2 \\
4\eta ^{2} & 4\eta ^{2}\varepsilon & -4\eta ^{2}\varepsilon & -2\eta
^{2}\left( 1+\varepsilon \right) & -2\eta ^{2}\left( 1-\varepsilon \right)
\\
4\eta ^{2} & \varepsilon ^{2}\left( 1-\varepsilon \right) /2 & \varepsilon
^{2}\left( 1+\varepsilon \right) /2 & 2\eta ^{2}\left( 1-\varepsilon \right)
& 2\eta ^{2}\left( 1+\varepsilon \right) \\
4\eta ^{2} & -4\eta ^{2}\varepsilon & 4\eta ^{2}\varepsilon & -2\eta
^{2}\left( 1-\varepsilon \right) & -2\eta ^{2}\left( 1+\varepsilon \right)
\\
4\eta ^{2} & \varepsilon \left( 1+\varepsilon \right) /2 & \varepsilon
\left( 1-\varepsilon \right) /2 & 2\eta ^{2}\left( 1+\varepsilon \right) &
2\eta ^{2}\left( 1-\varepsilon \right) \\
0 & -\eta \varepsilon \left( 1-\varepsilon \right) & -\eta \varepsilon
\left( 1+\varepsilon \right) & -\eta \left( 1-\varepsilon \right) ^{2}/2 &
\eta \left( 1+\varepsilon \right) ^{2}/2 \\
0 & \eta \varepsilon \left( 1+\varepsilon \right) & \eta \varepsilon \left(
1-\varepsilon \right) & -\eta \left( 1+\varepsilon \right) ^{2}/2 & \eta
\left( 1-\varepsilon \right) ^{2}/2
\end{array}
\right] \left[
\begin{array}{c}
\frac{\kappa }{\kappa +i\delta } \\
\\
\frac{\kappa }{\kappa +i\left( \delta -\Omega _{R}\right) } \\
\\
\frac{\kappa }{\kappa +i\left( \delta +\Omega _{R}\right) } \\
\\
\frac{\kappa }{\kappa +i\left( \delta -2\Omega _{R}\right) } \\
\\
\frac{\kappa }{\kappa +i\left( \delta +2\Omega _{R}\right) }
\end{array}
\right] ,
\end{equation}
with
\begin{equation}
\Omega _{R}=\frac{1}{2}\sqrt{\omega _{21}^{2}+8\Omega ^{2}},\;\;\;\eta =%
\frac{\Omega }{2\Omega _{R}},\;\;\;\varepsilon =\frac{\omega _{21}}{2\Omega
_{R}}.
\end{equation}
The first term of eq. (\ref{mast}) describes the coherent evolution of the
atom, the second terms the cavity-induced decay of the atom into the cavity
mode, and the remaining terms the atomic spontaneous emissions to the
background modes.

The equations of motion for the atomic variables take the form

\begin{eqnarray}
\dot{\rho}_{11} &=&-\gamma \rho _{11}-i\Omega \left( \rho _{01}-\rho
_{10}\right)  \nonumber \\
&&-\frac{\gamma _{c}}{2}\left( \beta _{0}\rho _{01}+\beta _{0}^{\ast }\rho
_{10}+\beta _{4}\rho _{11}+\beta _{4}^{\ast }\rho _{11}+\beta _{6}\rho
_{21}+\beta _{6}^{\ast }\rho _{12}\right) ,  \nonumber \\
\dot{\rho}_{22} &=&-\gamma \rho _{22}-i\Omega \left( \rho _{02}-\rho
_{20}\right)  \nonumber \\
&&-\frac{\gamma _{c}}{2}\left( \beta _{0}\rho _{02}+\beta _{0}^{\ast }\rho
_{20}+\beta _{4}\rho _{12}+\beta _{4}^{\ast }\rho _{21}+\beta _{6}\rho
_{22}+\beta _{6}^{\ast }\rho _{22}\right) ,  \nonumber \\
\dot{\rho}_{10} &=&-\frac{1}{2}\left( \gamma -i\omega _{21}\right) \rho
_{10}+i\Omega \left( \rho _{11}-\rho _{00}+\rho _{12}\right)  \nonumber \\
&&-\frac{\gamma _{c}}{2}\left[ \beta _{0}\rho _{00}-\beta _{1}\left( \rho
_{11}+\rho _{12}\right) -\beta _{3}\left( \rho _{01}+\rho _{02}\right)
+\beta _{4}\rho _{10}+\beta _{6}\rho _{20}-\beta _{8}\left( \rho _{21}+\rho
_{22}\right) \right] ,  \nonumber \\
\dot{\rho}_{20} &=&-\frac{1}{2}\left( \gamma +i\omega _{21}\right) \rho
_{20}+i\Omega \left( \rho _{22}-\rho _{00}+\rho _{21}\right)  \nonumber \\
&&-\frac{\gamma _{c}}{2}\left[ \beta _{0}\rho _{00}-\beta _{2}\left( \rho
_{21}+\rho _{22}\right) -\beta _{5}\left( \rho _{01}+\rho _{02}\right)
+\beta _{4}\rho _{10}+\beta _{6}\rho _{20}-\beta _{7}\left( \rho _{11}+\rho
_{12}\right) \right] ,  \nonumber \\
\dot{\rho}_{21} &=&-\left( \gamma +i\omega _{21}\right) \rho _{21}-i\Omega
\left( \rho _{01}-\rho _{20}\right)  \nonumber \\
&&-\frac{\gamma _{c}}{2}\left( \beta _{0}\rho _{01}+\beta _{0}^{\ast }\rho
_{20}+\beta _{4}\rho _{11}+\beta _{4}^{\ast }\rho _{21}+\beta _{6}\rho
_{21}+\beta _{6}^{\ast }\rho _{22}\right) ,  \label{dm}
\end{eqnarray}

\section{Steady-state populations}

\subsection{The bare states.}

We have numerically solved the equations (\ref{dm}) for the populations in
the steady state. In all our numerical plots we assume the values $\gamma
=1,g=20$ and $\kappa =100,$ so that the condition (\ref{bc}) is satisfied.
All the frequencies are also measured in units of $\gamma .$

In Fig. 2 we plot the bare state populations as a function of the
cavity-laser detuning $\delta .$\ We first consider the case where $\omega
_{21}=10,$ taking in frame (a) $\Omega =4;$ in frame (b), $\Omega =10;$ and
in frame (c), $\Omega =100$. For the smallest value of $\Omega ,$ all three
populations tend to roughly the same value for large detunings, but for
small detunings, a resonant effect is evident around $\delta =0:$\ the
population in the ground state passes through a maximum, whilst the
population in the two excited states exhibits a minimum. The behaviour in
frame (b) is qualitatively different: here, the population resides mainly in
the ground state, but this population now shows a dip as $\delta $ passes
through zero. The excited state populations show a minimum and a maximum
close to the origin. There is a very tiny amount of population inversion for
state $\left| 2\right\rangle $ over state $\left| 0\right\rangle $ for a
very short range of negative detunings close to zero, but the effect is
unimportant. As the value of $\Omega $ is increased to $\Omega =100$ in
frame (c), keeping $\omega _{21}=10,$ the populations show a flatter
behaviour, with the ground state population tending to the value 0.5 for
large $\delta ,$ and that of the excited states to the value 0.25. The
minimum in the ground state population at $\delta =0$ in frame (b) has been
replaced by a very shallow maximum in frame (c). It is clear that the most
interesting behaviour arises for $\Omega \sim \omega _{21}.$

In the next three frames we assume a larger value for the excited state
splitting, $\omega _{21}=200.$ For the lowest value of $\Omega $ considered,
$\Omega =100,$ the behaviour in frame (d) is qualitatively different to that
shown in the first three frames. Now all three bare populations tend towards
the same value for large $\delta .$ The ground state population is almost
flat, but the excited states possess pronounced maxima and minima, as well
as the suggestion of further structure. It is clear that, for appropriate
detunings, a large population inversion between the excited and ground
states can be achieved. When $\Omega $ is increased to $\Omega =200,$ as in
frame (e), the ground state population begins to increase, decreasing the
population inversion obtainable. Also structure at four different
frequencies becomes evident. (The reasons for this will become apparent in
the following subsection.) The trend continues in the final frame, where $%
\Omega =300,$ and there is no population inversion. For still larger values
of $\Omega $ (not shown), the ground state population tends to flatten
around the value 0.5, and the excited state populations around the value
0.25, as in frame (c) but with different detailed structure.

\subsection{The dressed states.}

To study the modification of the atomic populations due to the presence of
the cavity, we work in the semiclassical dressed-state representation. The
dressed states, defined by the eigenvalue equation, $H_{A}|\alpha \rangle
=\lambda _{\alpha }|\alpha \rangle $, are of the form \cite{s12}
\begin{eqnarray}
&&|a\rangle =\frac{1}{2}[-(1-\varepsilon )|2\rangle -(1+\varepsilon
)|1\rangle +4\eta |0\rangle ],  \nonumber \\
&&|b\rangle =-2\eta |2\rangle +2\eta |1\rangle +\varepsilon |0\rangle ,
\nonumber \\
&&|c\rangle =\frac{1}{2}[(1+\varepsilon )|2\rangle +(1-\varepsilon
)|1\rangle +4\eta |0\rangle ],  \label{ds}
\end{eqnarray}
and the corresponding energies are
\begin{equation}
\lambda _{a}=-\Omega _{R},\;\;\;\lambda _{b}=0,\;\;\;\lambda _{c}=\Omega
_{R}.
\end{equation}

Within the secular approximation, the equations of motion for the
populations in the dressed states may be cast into the rate equation form

\begin{eqnarray}
\dot{\rho}_{aa} &=&-(R_{ab}+R_{ac})\rho _{aa}+R_{ca}\rho _{cc}+R_{ba}\rho
_{bb}  \nonumber \\
\dot{\rho}_{bb} &=&-(R_{bc}+R_{ba})\rho _{bb}+R_{cb}\rho _{cc}+R_{ab}\rho
_{aa}  \nonumber \\
\dot{\rho}_{cc} &=&-(R_{ca}+R_{cb})\rho _{cc}+R_{ac}\rho _{aa}+R_{bc}\rho
_{bb}  \label{requ}
\end{eqnarray}
where $R_{\alpha \beta }\,(\alpha ,\,\beta =a,\,b,\,c)$ represents the
atomic transition rate from the substate $|\alpha \rangle $ of one
dressed-state triplet to the substate $|\beta \rangle $ of the dressed-state
triplet below, as depicted in Fig. 3. In the high field limit, that is, when
the effective Rabi frequency is much greater than all the relaxation rates, $%
\Omega _{R}\gg \gamma ,\,\gamma _{c}$, the coupling between atomic density
matrix elements $\rho _{\alpha \beta }$ associated with the various
frequencies may be omitted to $O(\gamma /\Omega _{R})$ and $O(\gamma
_{c}/\Omega _{R})$, and the transition rates may be expressed as
\begin{eqnarray}
&&R_{bc}=R_{ba}=\frac{\gamma }{2}(1-\varepsilon ^{2})^{2},  \nonumber \\
&&R_{ab}=\frac{\gamma }{2}(1+\varepsilon ^{2})\varepsilon ^{2}+\gamma
_{c}\varepsilon ^{2}{\cal R}(\Omega _{R}),  \nonumber \\
&&R_{cb}=\frac{\gamma }{2}(1+\varepsilon ^{2})\varepsilon ^{2}+\gamma
_{c}\varepsilon ^{2}{\cal R}(-\Omega _{R}),  \nonumber \\
&&R_{ac}=\frac{\gamma }{4}(1-\varepsilon ^{4})+\gamma _{c}4\eta ^{2}{\cal R}%
(2\Omega _{R}),  \nonumber \\
&&R_{ca}=\frac{\gamma }{4}(1-\varepsilon ^{4})+\gamma _{c}4\eta ^{2}{\cal R}%
(-2\Omega _{R}),  \label{trar}
\end{eqnarray}
where ${\cal R}(\pm x)=\kappa ^{2}/\left[ \kappa ^{2}+(\delta \pm x)^{2}%
\right] .$ The above expressions for the transition rates will provide the
basis for physical explanation of the effects to be described.

The function ${\cal R}(\pm 2\Omega _{R})$ introduces resonances at $\delta
=\pm 2\Omega _{R}$ into the rates $R_{ac}$ and $R_{ca}$ with a strength
proportional to $4\Omega ^{2}.$ This appears to be more dominant than the
resonances at $\delta =\pm \Omega _{R}$ which occur in $R_{cb}$ and $R_{ab}$
with a strength proportional to $\omega _{21}^{2}.$ Resonances at $\delta
=\pm 2\Omega _{R}$ can be clearly seen in frames (b) to (f) of Fig. 2,
whereas resonances at $\delta =\pm \Omega _{R}$ are only apparent in frames
(c) -- (f).

The equations (\ref{trar}) show that in the presence of the cavity, the
transition rates $R_{ab}$, $R_{ac}$, $R_{cb}$ and $R_{ca}$ are strongly
dependent on the cavity frequency, but $R_{bc}$ and $R_{ba}$ are only
related to the spontaneous emission rate. This is because for the system
considered here, a V-type three-level atom interacting with the cavity mode,
there exist nine double-channels for the atomic transition from the dressed
states $|i\rangle $ to $|j\rangle $, which originate from the atomic bare
state transitions $|1\rangle $ to $|0\rangle $ and $|2\rangle $ to $%
|0\rangle $ respectively. Constructive or destructive interference happens
within every double-channel. But the two double-channels for the transitions
$|b\rangle \rightarrow |c\rangle $ and $|b\rangle \rightarrow |a\rangle $
are completely destructive for the cavity mode---that is, the transitions
from $|b\rangle $ to $|c\rangle $ and from $|b\rangle $ to $|a\rangle $
never result in the emission of a photon into the cavity mode. So the atomic
transition rates $R_{bc}$ and $R_{ba}$ are only dependent on the spontaneous
emission rate and independent of the cavity frequency. As the other seven
double-channels are still open to the cavity mode, this means that for the
other transitions from $|i\rangle $ to $|j\rangle $, a cavity photon can be
generated, but it will be very strongly damped under the bad-cavity
assumption. Therefore, besides the terms describing atomic photon emission
into the background, there occur additional terms dependent on the cavity
frequency in the transition rates $R_{ab}$, $R_{ac}$, $R_{cb}$ and $R_{ca}$.

The steady-state dressed populations are found from equations (\ref{requ})
to be

\begin{eqnarray}
&&\rho _{aa}=\frac{R_{ba}(R_{ca}+R_{cb}+R_{bc})+R_{bc}(R_{ca}-R_{ba})}{%
(R_{ab}+R_{ac}+R_{ba})(R_{ca}+R_{cb}+R_{bc})-(R_{ca}-R_{ba})(R_{ac}-R_{bc})},
\nonumber \\
&&\rho _{cc}=\frac{R_{ba}(R_{ac}-R_{bc})+R_{bc}(R_{ab}+R_{ac}+R_{ba})}{%
(R_{ab}+R_{ac}+R_{ba})(R_{ca}+R_{cb}+R_{bc})-(R_{ca}-R_{ba})(R_{ac}-R_{bc})},
\nonumber \\
&&\rho _{bb}=1-\rho _{aa}-\rho _{cc}.  \label{dpops}
\end{eqnarray}
The dressed state populations are plotted as a function of the detuning in
Fig. 4, for the same parameter values employed in Fig. 3. These plots have
been obtained by a numerical solution of equations (\ref{dm}), but we have
found that the equations (\ref{dpops}) provide an excellent approximation in
the strong field limit.

It is evident that if the excited levels of the V-atom are degenerate ($%
\varepsilon =0$) or nearly degenerate ($\varepsilon \simeq 0$), the
transitions $|a\rangle \rightarrow |b\rangle $ and $|c\rangle \rightarrow
|b\rangle $ are turned off (as $R_{ab},R_{cb}\sim \varepsilon ^{2}$),
whereas the rate of transitions out of the dressed state $|b\rangle $, $%
R_{ba}+R_{bc},$ is nonzero. There is thus no steady-state population in the
dressed state $|b\rangle $, only in the dressed states $|a\rangle $ and $%
|c\rangle $. In general, it follows that in the regime $\Omega ^{2}\gg
\omega _{21}^{2},$ the population in dressed state $\left| b\right\rangle $
is very small. Furthermore, if the cavity frequency is tuned to $\delta
=-2\Omega _{R}$, the dressed state transition from $|a\rangle $ to $%
|c\rangle $ is resonantly enhanced ($R_{ac}\simeq \gamma /4+\gamma _{c}4\eta
^{2}$), while the reverse transition is suppressed, $R_{ca}\simeq \gamma /4$%
, and thus more population will be accumulated into the state $|c\rangle $.
For $\delta =2\Omega _{R}$, there exists a greater population in the dressed
state $|a\rangle $. An example is shown in frame (c) of Fig. 4, where we
take $\omega _{21}=10,\Omega =100.$ (The population in $\left|
b\right\rangle $ is very close to zero.) In frame (f), where we are only
beginning to approach the limit $\Omega ^{2}\gg \omega _{21}^{2},$ the
behaviour is similar, but now there is a small but significant population in
state $\left| b\right\rangle .$

However, in the opposite limit, where the excited level splitting is much
greater than the Rabi frequency ($\omega _{21}^{2}\gg \Omega ^{2}$) the
transition rates $R_{ac},\,R_{ca},\,R_{ba}$ and $R_{bc}$ are very small, and
the transitions into the dressed state $|b\rangle $, represented by $R_{ab}$
and $R_{cb}$, dominate. Eventually, the population in the dressed state $%
|b\rangle $ approaches unity, and that in the states $|a\rangle $ and $%
|c\rangle $ is very small. In this case, the resonance features of the
dressed state populations regarding the cavity frequency are less
pronounced. We do not plot this case here, but the features can be seen
beginning to emerge in frame (a), and to a lesser extent, in frame (c). The
intermediate regime, where $\omega _{21}=\Omega ,$ is shown in frames (b)
and (e).

In general, the population distributions are strongly dependent on the
cavity frequency. For example, when $\delta =0,$ the cavity is tuned to
resonance with the driving field, we have $R_{ab}=R_{cb}$, $R_{ba}=R_{bc}$
and $R_{ac}=R_{ca}$. Consequently, there is no population difference between
the upper dressed state $|c\rangle $ and the lower one $|a\rangle $.
Moreover, in the case $\Omega _{R}\gg \kappa $, the distribution is same as
in free space \cite{s12}.

When the cavity frequency satisfies $\delta =-\Omega _{R}$, describing
resonance with the dressed state transition $|a\rangle \rightarrow |b\rangle
$, the rate of this transition $R_{ab}$ is greatly enhanced. Also the rate
of downward atomic transitions from $|a\rangle $ to $|c\rangle $ is larger
than the transition rate $|c\rangle \rightarrow |a\rangle $, i.e., $%
R_{ac}>R_{ca}$. As a result, the population in the dressed state $|c\rangle $
is greater than that in the state $|a\rangle ,$ $(\rho _{cc}>\rho _{aa}).$

A similar analysis shows that the population, $\rho _{cc}$, of the state $%
|c\rangle $ is also greater than the population, $\rho _{aa}$, of the
dressed state $|a\rangle $ if the cavity is tuned to $\delta =-2\Omega _{R}$%
. The opposite conclusions hold when $\delta =\Omega _{R}$ and $2\Omega _{R}$%
, as is clearly demonstrated in Figs. 4(c) -- 4(f) for different values of
the Rabi frequency $\Omega $.

If the cavity is tuned to resonance with the driving field $(\delta =0),$
then $R_{ab}=R_{cb},$ $R_{ca}=R_{ac}$ and $R_{bc}=R_{ba}$. From Fig. 2 we
see that the population distribution between the dressed states $|c\rangle $
and $|a\rangle $ is balanced, which is similar to the case in the absence of
cavity \cite{s12}. In the presence of the cavity, the transition rates $%
R_{ab}$ and $R_{cb}\ $from the dressed states $|a\rangle $ and $|c\rangle $
to the dressed state $|b\rangle $ are faster than those in the absence of
the cavity, whilst the reverse rates $R_{bc}$ and $R_{ba}$ remain unchanged
because of the closing of the atomic transitions $|b\rangle \rightarrow
|c\rangle $ and $|b\rangle \rightarrow |a\rangle $ for the cavity mode due
to destructive interference. The populations $\rho _{cc}$ and $\rho _{aa}$
are decreased and $\rho _{bb}$ is increased by comparison with the case in
the absence of the cavity.

However, when the cavity frequency is tuned to satisfy $\delta =-\Omega _{R}$%
, the atomic transition from $|a\rangle $ to $|b\rangle $, which occurs at
frequency $\omega _{L}-\Omega _{R},$ is resonant with the cavity. The
interaction of the atom with the privileged cavity mode is enhanced, and the
atom predominately (since $g\gg \gamma $) emits a photon into the cavity
mode, characterized by the $\varepsilon ^{2}\gamma _{c}$ term in $R_{ab}$,
besides radiating a photon into the background, represented by the term $%
\Gamma _{0}$ in $R_{ab}$. By contrast, the other transitions describing
atomic emission of a photon into the cavity mode are far off resonance with
the cavity (since $\Omega _{R}\gg \kappa $), so the atom can only radiate a
photon into the background. As a consequence, the symmetry of the
transitions from $|c\rangle $ to $|b\rangle \;$and from $|a\rangle $ to $%
|b\rangle $ is broken, which results in an enhanced population in the
dressed state $|c\rangle $ over that in $|a\rangle $.

When the cavity frequency is tuned to satisfy $\delta =-2\Omega _{R}$, the
atomic transition rate $R_{ac}$ at the cavity frequency $\omega _{L}-2\Omega
_{R}$ is resonantly enhanced, and the other three rates $R_{ab}$, $R_{cb}$
and $R_{ca}$ are decreased. In this case, if $\gamma _{c}\gg \gamma \left(
1+\varepsilon ^{2}\right) /\left( 1-\varepsilon ^{2}\right) $, the other
five transition rates are much smaller than $R_{ca}$. Thus any population in
the dressed state $|a\rangle ,$ resulting from transitions from $|b\rangle $
and $|c\rangle $ to $|a\rangle ,$ will return to $|c\rangle $ very quickly
because $R_{ac}$ is much greater than the other transition rates. So it
seems that the atom is trapped in the two dressed states $|b\rangle $ and $%
|c\rangle $, and the population in the state $|a\rangle $ approaches zero.
Similar explanations can be adopted for the cases $\delta =\Omega _{R}$ and $%
\delta =2\Omega _{R}$.

\section{Resonance fluorescence spectrum}

The spectrum of the atomic fluorescence emission emitted from the side of
the cavity is proportional to the Fourier transform of the steady-state
correlation function $\lim_{t\rightarrow \infty }\langle \vec{E}^{(-)}(\vec{r%
},t+\tau )\cdot \vec{E}^{(+)}(\vec{r},t)\rangle $, where $\vec{E}^{(\pm )}(%
\vec{r},t)$ are the positive and negative frequency parts of the radiation
field in the far zone, which consists of a free-field operator and a source
field that is proportional to the atomic polarization operator. The
fluorescence spectrum is given by
\begin{equation}
\Lambda (\omega )=%
\mathop{\rm Re}%
\int\limits_{0}^{\infty }[\langle A_{20}(t+\tau ),A_{02}(t)\rangle +\langle
A_{10}(t+\tau ),A_{01}(t)\rangle ]_{t\rightarrow \infty }e^{-i\omega \tau
}d\tau .
\end{equation}
Because we have assumed that the atomic dipole moment elements $\vec{d}_{20}$
and $\vec{d}_{10}$ are perpendicular to each other, the two correlation
functions $\langle A_{10}(t+\tau ),A_{02}(t)\rangle $ and $\langle
A_{20}(t+\tau ),A_{01}(t)\rangle $ make no contribution to the fluorescence
emission spectrum. Using the quantum regression theorem, the spectrum may be
expressed as
\begin{equation}
\Lambda (\omega )=%
\mathop{\rm Re}%
\left[ \tilde{F}_{01}\left( z\right) +\tilde{G}_{02}\left( z\right) -\left(
\left| \rho _{01}\left( \infty \right) \right| ^{2}+\left| \rho _{02}\left(
\infty \right) \right| ^{2}\right) /z\right] _{z=i\omega }  \label{qrt1}
\end{equation}
where $\tilde{X}\left( z\right) $ denotes the Laplace transform of $X\left(
\tau \right) $, and
\begin{equation}
F\left( \tau \right) =U^{\dag }\left( \tau \right) \left[ \left|
0\right\rangle \left\langle 1\right| \rho \left( \infty \right) \right]
U\left( \tau \right) \,\text{and }G\left( \tau \right) =U^{\dag }\left( \tau
\right) \left[ \left| 0\right\rangle \left\langle 2\right| \rho \left(
\infty \right) \right] U\left( \tau \right)   \label{qrt2}
\end{equation}
with $U\left( t\right) $ the time development operator, obey the same
equation of motion as $\rho \left( \tau \right) $ but with the different
initial conditions implied by the definitions (\ref{qrt2}).

We calculate the resonance fluorescence spectra numerically from the
equations obtained in Section II. However, the dressed atom approach
provides a convenient way of interpreting the results so obtained, at least
in the strong field limit, and so we develop this approach here. In terms of
the dressed states (\ref{ds}), we have, for example,
\begin{equation}
\tilde{F}_{01}\left( z\right) =\sum_{\alpha ,\beta =a,b,c}\left\langle
0|\alpha \right\rangle \tilde{F}_{\alpha \beta }\left( z\right) \left\langle
\beta |1\right\rangle  \label{F}
\end{equation}
with the initial condition
\begin{equation}
F_{\alpha \beta }\left( 0\right) =\sum_{\gamma }\left\langle \alpha
|0\right\rangle \rho _{\gamma \beta }\left( \infty \right) \left\langle
1|\beta \right\rangle .
\end{equation}

In the high field limit, $\Omega _{R}\gg \gamma ,\gamma _{c}$, which is the
condition for the secular approximation to hold, the dressed energy levels
are well separated, and one can associate the diagonal terms on the right
hand side of equation (\ref{F}), and those in the corresponding equation for
$\tilde{G}_{02}\left( z\right) ,$ with the resonance fluorescence line at
the centre of the spectrum,$\ \omega =\omega _{L}$ (see Fig. 3). In a
similar way, the elements $\tilde{F}_{ab},\tilde{G}_{ab},\tilde{F}_{bc}$ and
$\tilde{G}_{bc}$ are associated with the line at $\omega =\omega _{L}-\Omega
_{R}$, whilst the elements $\tilde{F}_{ba},\tilde{G}_{ba},\tilde{F}_{cbc}$
and $\tilde{G}_{cb}$ are associated with the line at $\omega =\omega
_{L}+\Omega _{R}.$ Finally, $\tilde{F}_{ac}$ and $\tilde{G}_{ac}$ are
associated with the line at $\omega =\omega _{L}+2\Omega _{R},$ and $\tilde{F%
}_{ca}$ and $\tilde{G}_{ca}$ with the line at $\omega =\omega _{L}-2\Omega
_{R}$

One can apply the secular approximation to simplify the equations of motion
for the atomic density matrix elements in the dressed state representation,
which are obtained according to eq. (\ref{mast}) as follows
\begin{eqnarray}
&&\dot{\rho}_{aa}=-\Gamma _{1a}\rho _{aa}+\Gamma _{2a}\rho _{cc}+R_{ba},
\nonumber \\
&&\dot{\rho}_{cc}=-\Gamma _{1b}\rho _{cc}+\Gamma _{2b}\rho _{aa}+R_{bc},
\nonumber \\
&&\dot{\rho}_{ab}=-\left( \Gamma _{3a}-i\Omega _{3}\right) \rho _{ab}+\Gamma
_{4}\rho _{bc},  \nonumber \\
&&\dot{\rho}_{bc}=-\left( \Gamma _{3b}-i\Omega _{4}\right) \rho _{bc}+\Gamma
_{4}\rho _{ab},  \nonumber \\
&&\dot{\rho}_{ac}=-\left( \Gamma _{5}-i\Omega _{5}\right) \rho _{ac},
\end{eqnarray}
with
\begin{eqnarray}
&&\Gamma _{1a}=\frac{\gamma }{4}(3-2\varepsilon ^{2}+3\varepsilon
^{4})+\gamma _{c}\left[ \varepsilon ^{2}{\cal R}(\Omega _{R})+4\eta ^{2}%
{\cal R}(2\Omega _{R})\right] ,  \nonumber \\
&&\Gamma _{1b}=\frac{\gamma }{4}(3-2\varepsilon ^{2}+3\varepsilon
^{4})+\gamma _{c}\left[ \varepsilon ^{2}{\cal R}(-\Omega _{R})+4\eta ^{2}%
{\cal R}(-2\Omega _{R})\right] ,  \nonumber \\
&&\Gamma _{2a}=\frac{\gamma }{2}\left( 1-\varepsilon ^{2}\right) \left(
3\varepsilon ^{2}-1\right) +\gamma _{c}4\eta ^{2}{\cal R}(-2\Omega _{R}),
\nonumber \\
&&\Gamma _{2b}=\frac{\gamma }{2}\left( 1-\varepsilon ^{2}\right) \left(
3\varepsilon ^{2}-1\right) +\gamma _{c}4\eta ^{2}{\cal R}(2\Omega _{R}),
\nonumber \\
&&\Gamma _{3a}=\frac{\gamma }{4}(3+\varepsilon ^{2}-2\varepsilon ^{4})+\frac{%
\gamma _{c}}{2}\left[ 4\eta ^{2}{\cal R}(0)+\varepsilon ^{2}{\cal R}(\Omega
_{R})+4\eta ^{2}{\cal R}(2\Omega _{R})\right] ,  \nonumber \\
&&\Gamma _{3b}=\frac{\gamma }{4}(3+\varepsilon ^{2}-2\varepsilon ^{4})+\frac{%
\gamma _{c}}{2}\left[ 4\eta ^{2}{\cal R}(0)+\varepsilon ^{2}{\cal R}(-\Omega
_{R})+4\eta ^{2}{\cal R}(-2\Omega _{R})\right] ,  \nonumber \\
&&\Gamma _{4}=-\frac{\gamma }{2}\varepsilon ^{2}(1-\varepsilon ^{2}),
\nonumber \\
&&\Gamma _{5}=\frac{\gamma }{4}(3+\varepsilon ^{4})+\frac{\gamma _{c}}{2}%
\left\{ 16\eta ^{2}{\cal R}(0)+\varepsilon ^{2}\left[ {\cal R}(\Omega _{R})+%
{\cal R}(-\Omega _{R})\right] +4\eta ^{2}\left[ {\cal R}(2\Omega _{R})+4\eta
^{2}{\cal R}(-2\Omega _{R})\right] \right\} ,  \nonumber \\
&&\Omega _{3}=\Omega _{R}+\frac{\gamma _{c}}{2}\left[ 4\eta ^{2}{\cal I}%
(0)+\varepsilon ^{2}{\cal I}(\Omega _{R})+4\eta ^{2}{\cal I}(2\Omega _{R})%
\right] ,  \nonumber \\
&&\Omega _{4}=\Omega _{R}+\frac{\gamma _{c}}{2}\left[ 4\eta ^{2}{\cal I}%
(0)+\varepsilon ^{2}{\cal I}(-\Omega _{R})+4\eta ^{2}{\cal I}(-2\Omega _{R})%
\right] ,  \nonumber \\
&&\Omega _{5}=2\Omega _{R}+\frac{\gamma _{c}}{2}\{\varepsilon ^{2}\left[
{\cal I}(\Omega _{R})-{\cal I}(-\Omega _{R})\right] +4\eta ^{2}\left[ {\cal I%
}(2\Omega _{R})-{\cal I}(-2\Omega _{R})\right] \},  \label{drates}
\end{eqnarray}
where ${\cal I}(\pm x)=\kappa (\delta \pm x)/\left[ \kappa ^{2}+(\delta \pm
x)^{2}\right] $, $\Gamma _{i}$ is the decay rate in the dressed state
representation, which is dependent on the cavity frequency, while $\Omega
_{R}-\Omega _{3,4}$ and $2\Omega _{R}-\Omega _{5}$ are the cavity-induced
level shifts. In the bad cavity and high field limits, the shifts are
negligibly small.

In the dressed state representation, the underlying physical processes are
very transparent. As argued in the paragraph following equation (\ref{F}),
the downward transitions between the same dressed states of two adjacent
dressed-state triplets give rise to the central component of the
fluorescence spectrum, i.e.,
\begin{equation}
\Lambda _{0}(\omega )=%
\mathop{\rm Re}%
\left[ \frac{N_{0}(z)}{\left( z+\Gamma _{1a}\right) \left( z+\Gamma
_{1b}\right) -\Gamma _{2a}\Gamma _{2b}}\right] _{z=i\omega },  \label{center}
\end{equation}
with
\begin{eqnarray}
N_{0}(z) &=&4\eta ^{2}\left( 2z+\Gamma _{1a}+\Gamma _{1b}-\Gamma
_{2a}-\Gamma _{2b}\right) \rho _{aa}\rho _{cc}-2\eta ^{2}\left(
1-9\varepsilon ^{2}\right) \left[ \Gamma _{2a}\rho _{cc}+\Gamma _{2b}\rho
_{aa}\right] \rho _{bb}  \nonumber \\
&&+2\eta ^{2}\left( 1+9\varepsilon ^{2}\right) \left[ \left( z+\Gamma
_{1a}\right) \rho _{cc}+\left( z+\Gamma _{1b}\right) \rho _{aa}\right] \rho
_{bb}.
\end{eqnarray}
This spectral component consists of two Lorentzians with linewidths $2\gamma
_{0}^{\pm }=\left( \Gamma _{1a}+\Gamma _{1b}\right) \pm \sqrt{\left( \Gamma
_{1a}-\Gamma _{1b}\right) ^{2}+4\Gamma _{2a}\Gamma _{2b}}$.

However, the downward transitions $|a\rangle \rightarrow |b\rangle \,$and $%
|b\rangle \rightarrow |c\rangle $ from one dressed-state triplet to the next
triplet lead to the lower-frequency inner sideband, yielding an expression
of the form

\begin{equation}
\Lambda _{1}(\omega )=%
\mathop{\rm Re}%
\left[ \frac{4\eta ^{2}\left[ 8\eta ^{2}\left( z+\Gamma _{3a}+i\Omega
_{3}\right) -\varepsilon ^{2}\Gamma _{4}\right] \rho _{bb}+\frac{1}{2}%
\varepsilon ^{2}\left[ \left( 1+\varepsilon ^{2}\right) \left( z+\Gamma
_{3b}+i\Omega _{4}\right) -8\eta ^{2}\Gamma _{4}\right] \rho _{aa}}{\left(
z+\Gamma _{3a}+i\Omega _{3}\right) \left( z+\Gamma _{3b}+i\Omega _{4}\right)
-\Gamma _{4}^{2}}\right] _{z=i\omega },  \label{inner1}
\end{equation}
whilst the transitions $|b\rangle \rightarrow |a\rangle $ and $|c\rangle
\rightarrow |b\rangle $ between two near-lying dressed-state triplets result
in the higher-frequency inner sideband,

\begin{equation}
\Lambda _{2}(\omega )=%
\mathop{\rm Re}%
\left[ \frac{4\eta ^{2}\left[ 8\eta ^{2}\left( z+\Gamma _{3b}-i\Omega
_{4}\right) -\varepsilon ^{2}\Gamma _{4}\right] \rho _{bb}+\frac{1}{2}%
\varepsilon ^{2}\left[ \left( 1+\varepsilon ^{2}\right) \left( z+\Gamma
_{3a}-i\Omega _{3}\right) -8\eta ^{2}\Gamma _{4}\right] \rho _{cc}}{\left(
z+\Gamma _{3a}-i\Omega _{3}\right) \left( z+\Gamma _{3b}-i\Omega _{4}\right)
-\Gamma _{4}^{2}}\right] _{z=i\omega }.  \label{inner2}
\end{equation}
Since the cavity-induced level shifts are negligible, $\Lambda _{1}$ will
display a single spectral line located at frequency $\omega _{L}-\Omega _{R}$%
, and $\Lambda _{2}$ a line at $\omega _{L}+\Omega _{R}$. It is evident that
the inner sidebands are also composed of two Lorentzians with linewidths $%
2\gamma _{1}^{\pm }=\left( \Gamma _{3a}+\Gamma _{3b}\right) \pm \sqrt{\left(
\Gamma _{3a}-\Gamma _{3b}\right) ^{2}+4\Gamma _{4}^{2}}$.

The final transitions, $|a\rangle \rightarrow |c\rangle $ and $|a\rangle
\rightarrow |c\rangle ,$ respectively generate the lower-frequency and
higher-frequency spectral lines of the outer sidebands, which are given by

\begin{eqnarray}
\Lambda _{3}(\omega ) &=&%
\mathop{\rm Re}%
\left[ \frac{2\eta ^{2}\left( 1+\varepsilon ^{2}\right) \rho _{aa}}{\left(
z+\Gamma _{5}+i\Omega _{5}\right) }\right] _{z=i\omega },  \label{out1} \\
\Lambda _{4}(\omega ) &=&%
\mathop{\rm Re}%
\left[ \frac{2\eta ^{2}\left( 1+\varepsilon ^{2}\right) \rho _{cc}}{\left(
z+\Gamma _{5}-i\Omega _{5}\right) }\right] _{z=i\omega }.  \label{out2}
\end{eqnarray}
The spectral lines are centred at frequencies $\omega _{L}\pm 2\Omega _{R}$,
respectively, and have width $2\Gamma _{5}$. In equations (\ref{center}) -- (%
\ref{out2}), the $\rho _{jj}$ which appear are the steady state
dressed state occupation probabilities, given by equations
(\ref{dpops})

We first consider the fluorescence spectrum of the atom with two degenerate
(or near-degenerate) excited states, where the population of the dressed
state $|b\rangle $ is negligible, as illustrated in Fig. 5, where $\omega
_{21}=10$ and $\Omega =100,$ and we consider different detunings. The
results may be understood from equations (\ref{center}) -- (\ref{out2}). It
is not difficult to see from eqs. (\ref{inner1}) and (\ref{inner2}) that the
inner sidebands $\Lambda _{1,2}$ will be very small, as $\rho _{bb}\simeq 0$
and $\varepsilon ^{2}\simeq 0$. Hence the central line and the outer
sidebands will dominate. This is evident in all the spectra of Fig. 5, but
particularly for frame (a) where the inner sidebands are hardly visible.
When the cavity frequency $\delta =0$, the spectrum is a symmetric,
Mollow-like triplet, whilst the lower-frequency outer sideband is enhanced
and the higher-frequency one suppressed when $\delta =\Omega _{R}$ and $%
2\Omega _{R}$. These features are similar to those of a laser-dressed
two-level atom coupled to such a cavity field \cite{s7}. The enhancement and
suppression of the sidebands is due to the cavity modification of the
transition rates $|a\rangle \rightarrow |c\rangle $ and $|c\rangle
\rightarrow |a\rangle $. The former decreases while the latter increases for
$\delta =\Omega _{R},\,2\Omega _{R}$. Therefore, the enhanced spectral line
of the outer sidebands is narrowed whilst the suppressed one is broadened.
For large values of the detuning, shown in frame (d), the spectrum reverts
to a symmetric form.

Next, we display the modified spectrum in the limit of $\omega _{21}\gg
\Omega $ in Fig. 6 where the values $\omega _{21}=200$ and $\Omega =50$ are
taken. Contrary to the spectra of Fig. 5, the inner sidebands are most
pronounced for the $\delta =0$ situation whereas the outer spectral lines
are almost invisible. Tuning the cavity frequency may also change the
spectral profile: the higher-frequency sideband is somewhat enhanced when $%
\delta =\Omega _{R}$ and $2\Omega _{R}$. As we discussed above, in the limit
of $\omega _{21}\gg \Omega $ the dressed state populations $\rho _{aa}$ and $%
\rho _{cc}$ are close to zero while $\rho _{bb}\simeq 1$, and all the
populations are barely dependent on the cavity frequency. It is obvious from
eqs. (\ref{inner1}) and (\ref{inner2}) that the $\Lambda _{1,2}$ are mainly
determined by the cavity-frequency-dependent decay rates, $\Gamma _{3a}$ and
$\Gamma _{3b}$. For the parameters of Fig. 6, we have $\Gamma _{3a}(\delta
=0)=\Gamma _{3a}(\delta =0)\simeq 2.51$, $\Gamma _{3a}(\delta =\Omega
_{R})\simeq 1.33,\,\Gamma _{3b}(\delta =\Omega _{R})\simeq 4.40$ and $\Gamma
_{3a}(\delta =2\Omega _{R})=0.93,\,\Gamma _{3b}(\delta =2\Omega _{R})\simeq
2.50$. Therefore, $\Lambda _{1}\left( \omega =-\Omega _{R}\right) <\Lambda
_{2}\left( \omega =\Omega _{R}\right) $, which implies that the
lower-frequency peak is lower than the higher-frequency sideband.

In Fig. 7, we take the values $\omega _{21}=200,\Omega =100,$ which
corresponds to the populations shown in frame (d) of Fig. 4. For $\delta =0,$
the bulk of the population is concentrated in state $\left| b\right\rangle ,$
and there is little population in states $|a\rangle \ $and $|c\rangle .$
Hence the outer sidebands are weak in amplitude, whilst the inner sidebands
and the central peak are pronounced. Tuning the laser to $\delta =\Omega _{R}
$ or $\delta =2\Omega _{R}$ increases the relative amplitude of the high
frequency or low frequency inner sideband respectively. As $\delta $ is
increased through $\delta =\Omega _{R}$ to $\delta =2\Omega _{R},$ the
population in state $\left| a\right\rangle $ increases, whilst that in state
$\left| c\right\rangle $ decreases. The low frequency outer sideband
consequently increases in amplitude at the expence of the amplitude of the
high frequency outer sideband.

For the general case, by inspection of equations (\ref{drates}) for the
cavity-induced decay rates $\Gamma _{i}$, one finds in the limit $\Omega
_{R}\gg \kappa $ that if the cavity is tuned to resonance with the driving
laser $(\delta =0)$, the rates are the same as those in free space \cite{s12}%
, except for $\Gamma _{3}$ and $\Gamma _{5}$ being replaced by $\Gamma
_{3}+\gamma _{c}2\eta ^{2}$ and $\Gamma _{5}+\gamma _{c}8\eta ^{2}$
respectively. This reflects the fact that the cavity-induced spontaneous
emission rates can be greatly suppressed by increasing the Rabi frequency
\cite{s7}. As shown in Fig. 8(a) for $\omega _{21}=200$ and $\Omega =200$,
the inner and outer sidebands are broadened, whilst the central peak is
narrowed.

We illustrate the spectra for $\delta =\Omega _{R}$ and $2\Omega _{R}$ in
frames 8(b) and 8(c), respectively, where asymmetric spectral features are
exhibited. The previous explanations apply to this case as well, that is,
the enhancement and suppression of the spectral lines stems from the cavity
modification of the dressed state population distribution and decay rates.
For example, for $\delta =2\Omega _{R}$ the population in the dressed state $%
|a\rangle $ is much greater than those in the dressed states $|b\rangle $
and $|c\rangle $, as shown in Fig. 4(e). Accordingly, the lower-frequency
peaks are higher than their counterparts in the high frequency side.

\section{Absorption spectrum}

It is also natural to study the absorption spectrum of a weak, tunable probe
field transmitted through the system, as we would expect this to be largely
determined by the dressed states populations. The frequency $\omega $ of the
driving laser is kept fixed, and the absorption is measured as a function of
the frequency $\nu $ of the probe laser, measured from the laser frequency, $%
\omega .$ The absorption spectrum is given by
\begin{equation}
{\cal A}(\nu )=%
\mathop{\rm Re}%
\int\limits_{0}^{\infty }\left\{ \langle \left[ A_{20}(t+\tau ),A_{02}(t)%
\right] \rangle +\langle \left[ A_{10}(t+\tau ),A_{01}(t)\right] \rangle
\right\} _{t\rightarrow \infty }e^{i\nu \tau }d\tau .
\end{equation}
We present only a brief discussion of this phenomenon here. As for the
resonance fluorescence spectrum, we may use the quantum regression theorem
to express the probe spectrum as
\begin{equation}
{\cal A}(\omega )=%
\mathop{\rm Re}%
\left[ \tilde{F}_{01}\left( z\right) +\tilde{G}_{02}\left( z\right) -\tilde{F%
}_{10}^{\prime }\left( z\right) -\tilde{G}_{20}^{\prime }\left( z\right) %
\right] _{z=i\nu }
\end{equation}
where
\begin{equation}
F^{\prime }\left( \tau \right) =U^{\dag }\left( \tau \right) \left[ \rho
\left( \infty \right) \left| 1\right\rangle \left\langle 0\right| \right]
U\left( \tau \right) \,\text{and }G^{\prime }\left( \tau \right) =U^{\dag
}\left( \tau \right) \left[ \rho \left( \infty \right) \left| 2\right\rangle
\left\langle 0\right| \right] U\left( \tau \right) .
\end{equation}
It is possible to use the secular approximation to obtain expressions for
the spectral components, as was done in the previous section. However, for
brevity, we concentrate on a qualitative discussion here.

Figure 9, frames (a)--(c), shows the spectra for $\omega _{21}=10,\Omega
=100,(\Omega _{R}=141.5)$ when most of the population resides in the dressed
states $\left| a\right\rangle $ and $\left| c\right\rangle ,$ and the
population in state $\left| b\right\rangle $ is practically zero, although
the latter does show a very small maximum for zero detuning, $\delta =0.$
The case $\delta =0$ is shown in frame (a). Because there is no population
in state $\left| b\right\rangle ,$ the absorption spectrum is largely
determined by the transitions between the outer two levels from $\left|
c\right\rangle $ to $\left| a\right\rangle ,$ at frequency $\nu \simeq
2\Omega _{R},$ and by the transitions between the inner two levels $\left|
a\right\rangle $ to $\left| c\right\rangle ,$ at frequency $\nu \simeq
-2\Omega _{R}.$ For example, from Figure 4, following the arguments of
Cohen-Tannoudji and Reynaud \cite{co-re78}, one can write down an
approximate expression for the weight of the line at $\nu \simeq -2\Omega
_{R}$ as
\begin{equation}
w_{a}\left( -2\Omega _{R}\right) \simeq P_{a}R_{ac}-P_{c}R_{ca},
\end{equation}
a negative value for $w$ representing amplification, and a positive value
absorption. It is easily seen that
\begin{equation}
w_{a}\left( +2\Omega _{R}\right) \simeq
P_{c}R_{ca}-P_{a}R_{ac},=-w_{a}\left( -2\Omega _{R}\right)
\end{equation}
Since the populations in levels $\left| a\right\rangle $ and $\left|
c\right\rangle $ are equal when $\delta =0,$ the difference between
absorption and amplification is determined by the value of the rates.

A similar argument gives
\begin{equation}
w_{a}\left( -\Omega _{R}\right) \simeq
P_{c}R_{cb}+P_{b}R_{ba}-P_{b}R_{bc}-P_{a}R_{ab}\simeq
P_{c}R_{cb}-P_{a}R_{ab}.
\end{equation}
Using the expressions (\ref{trar}) for the rates, we find
\begin{equation}
\left| \frac{w_{a}\left( -\Omega _{R}\right) }{w_{a}\left( +2\Omega
_{R}\right) }\right| \simeq \frac{\omega _{21}^{2}}{4\Omega _{R}^{2}}\ll 1.
\end{equation}
These arguments explain the main features of the plot presented in frame
(a). In frame (b), we tune the driving laser to $\delta =\Omega _{R}$ which
has the effect of significantly increasing the population in $\left|
c\right\rangle $ and decreasing that in level $\left| a\right\rangle .$ The
magnitudes of the absorption at $\nu \simeq -2\Omega _{R},$ and the emission
at $\nu \simeq 2\Omega _{R},$ are thus greatly increased. The difference
between the populations $\left| a\right\rangle $ and $\left| c\right\rangle $
reaches a maximum for $\delta =\pm 2\Omega _{R}$ . This value of $\delta $
is assumed in frame (c), where the greatest absorption/emission occurs.

In frames (d)--(f), we change the parameters to $\omega _{21}=200,\Omega
=50\;\left( \Omega _{R}=122.5\right) .$ The populations for this case, whose
behaviour varies greatly from the previous case, are shown in frame (c) of
Fig. 4. Now we have a large population in state $\left| b\right\rangle ,$
and only small populations in states $\left| a\right\rangle $ and $\left|
c\right\rangle .$ The absorption spectra are thus quite different: the
strongest features occur at $\nu =\pm \Omega _{R},$ and have a strong
dispersive element, whereas the features at $\nu =\pm 2\Omega _{R}$ are
relatively insignificant. Tuning the driving laser to $\delta =\pm \Omega
_{R}$ or $\delta =\pm 2\Omega _{R}$ again has the effect of greatly
increasing the strengths of the absorptions/emissions.

The first three frames of Fig. 10 show the absorption spectra for $\omega
_{21}=200,\Omega =100\;\left( \Omega _{R}=173.2\right) $ and the second
three frames the absorption spectra for $\omega _{21}=200,\Omega
=200\;\left( \Omega _{R}=300\right) .$ In these two cases, changing the
detuning $\delta $ has a greater effect than in the cases of the previous
figure. This is particularly true in the last three frames, where it can be
seen that changing the detuning can cause a switch from strong absorption to
strong emission, and vice-versa.

\section{Conclusions}

We have investigated, in the bad cavity limit, the resonance fluorescence of
a V-type three-level atom strongly driven by a laser field and weakly
coupled to a cavity mode, and we have demonstrated how the fluorescence may
be controlled and manipulated by varying the cavity and Rabi frequencies. A
strong dependency of the bare and dressed state populations on the cavity
resonant frequency is shown. Population inversion in both the bare and the
dressed-state bases can be achieved for appropriate atom-cavity coupling
constants, cavity resonant frequency and high driving intensities. These
population inversions result from the enhanced atom-cavity interaction when
the cavity is tuned to resonance with the atomic dressed-state transition.
The resultant fluorescence spectrum is also strongly dependent on the cavity
frequency. When the cavity is in resonant with the driving field, the
spectrum is symmetric. Specifically, if the excited levels of the atom are
degenerate (or near-degenerate), a Mollow-like triplet is exhibited with
central line narrowing, but the spectrum has a two-peak structure if the
level splitting is much greater than the laser intensity. Otherwise, the
spectrum consists of five peaks.  When the cavity is tuned to resonance with
one of the spectral sidebands, some of the spectral lines may be enhanced
and others suppressed, and so the spectrum is asymmetric. Dynamical
line-narrowing and peak-suppression of certain spectral lines can be
achieved by increasing the laser intensity. We have also demonstrated that a
large degree of control and manipulation of the probe absorption spectrum
can be achieved by tuning the cavity.

\acknowledgments

JSP wishes to thank the Royal Society London for the financial support. This
work is supported by the Natural Science Foundation of China, and the United
Kingdom EPSRC.

\begin{figure}[tbp]
\caption{Configuration of a V-type three-level atom coupled to a single-mode
cavity and driven by a laser field.}
\label{fig1}
\end{figure}

\begin{figure}[tbp]
\caption{The bare populations of a V-type three-level atom coupled to a
single-mode cavity and driven by a laser field. The ground state population
is represented by a solid line, the first excited state by a dashed line,
and the second excited state by a dotted line. In all our figures, we
assume the values $\protect\gamma=1,g=20,\protect\kappa=100$. In the first
three frames here, we assume $\protect\omega_{21}=10$, and in the last three
$\protect\omega_{21}=100$. In frames (a), (b) and (c), we take $\Omega=4, 10$
and $100$ respectively, and in frames (d), (e) and (f) we take $\Omega=100,
200$ and $300$ respectively.}
\label{fig2}
\end{figure}

\begin{figure}[tbp]
\caption{Diagram of atomic dressed states and dressed-state transitions.}
\label{fig3}
\end{figure}

\begin{figure}[tbp]
\caption{The dressed-state populations, for the same parameters used in Fig.
2. The dressed state `b' is represented by a solid line, the dressed state `a'
by a dashed line, and the dressed state `c' by a dotted line. In the first
three frames, $\protect\omega_{21}=10$, and in the last three
$\protect\omega_{21}=100$. In frames (a), (b) and (c), $\Omega=4, 10$ and $100$
respectively, and in frames (d), (e) and (f), $\Omega=100, 200$ and $300$
respectively.}
\label{fig4}
\end{figure}

\begin{figure}[tbp]
\caption{The fluorescence spectrum for $\protect\omega_{21}=10,\, \Omega=100$%
, and (a), $\protect\delta=0$, (b), $\protect\delta=\Omega_R$, (c), $\protect%
\delta=2\Omega_R$ and (d), $\protect\delta=10\Omega_R$, respectively.}
\label{fig5}
\end{figure}

\begin{figure}[tbp]
\caption{The fluorescence spectrum for $\protect\omega_{21}=20,\, \Omega=50$%
, and (a), $\protect\delta=0$, (b), $\protect\delta=\Omega_R$, (c), $\protect%
\delta=2\Omega_R$ and (d), $\protect\delta=10\Omega_R$, respectively.}
\label{fig6}
\end{figure}

\begin{figure}[tbp]
\caption{The fluorescence spectrum for $\protect\omega_{21}=200,\, \Omega=100
$, and (a), $\protect\delta=0$, (b), $\protect\delta=\Omega_R$, (c), $%
\protect\delta=2\Omega_R$ and (d), $\protect\delta=10\Omega_R$, respectively.
}
\label{fig7}
\end{figure}

\begin{figure}[tbp]
\caption{The fluorescence spectrum for $\protect\omega_{21}=200,\, \Omega=200
$, and (a), $\protect\delta=0$, (b), $\protect\delta=\Omega_R$, (c), $%
\protect\delta=2\Omega_R$ and (d), $\protect\delta=10\Omega_R$, respectively.
}
\label{fig8}
\end{figure}

\begin{figure}[tbp]
\caption{The probe absorption spectrum. In the first three frames, $\protect%
\omega_{21}=10, \Omega=100$, and in the last three $\protect\omega_{21}=200,
\Omega=50$. In frames (a) and (d), $\protect\delta=0$, in frames (b) and (e), %
$\protect\delta=\Omega_R$, and in frames (c) and (f), $\protect\delta%
=2\Omega_R$, respectively.}
\label{fig9}
\end{figure}

\begin{figure}[tbp]
\caption{The probe absorption spectrum. In the first three frames, $\protect%
\omega_{21}=200, \Omega=100$, and in the last three $\protect\omega%
_{21}=200, \Omega=200$. In frames (a) and (d), $\protect\delta=0$, in frames
(b) and (e), $\protect\delta=\Omega_R$, and in frames (c) and (f), $\protect%
\delta=2\Omega_R$, respectively.}
\label{fig10}
\end{figure}

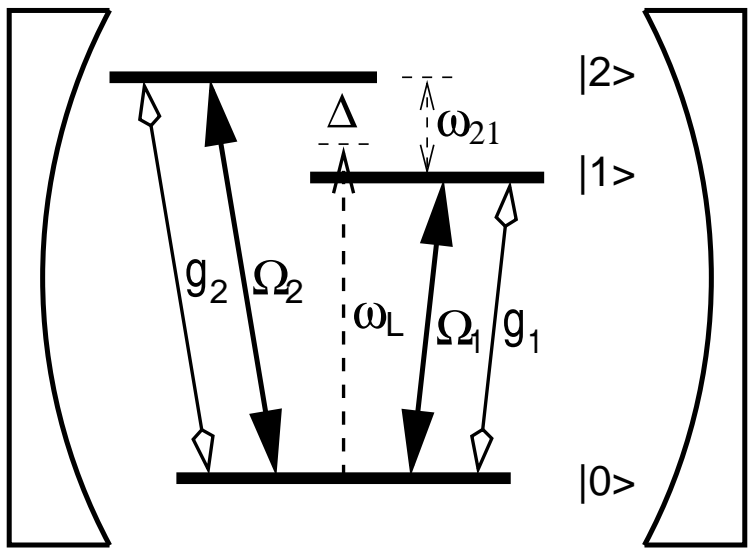
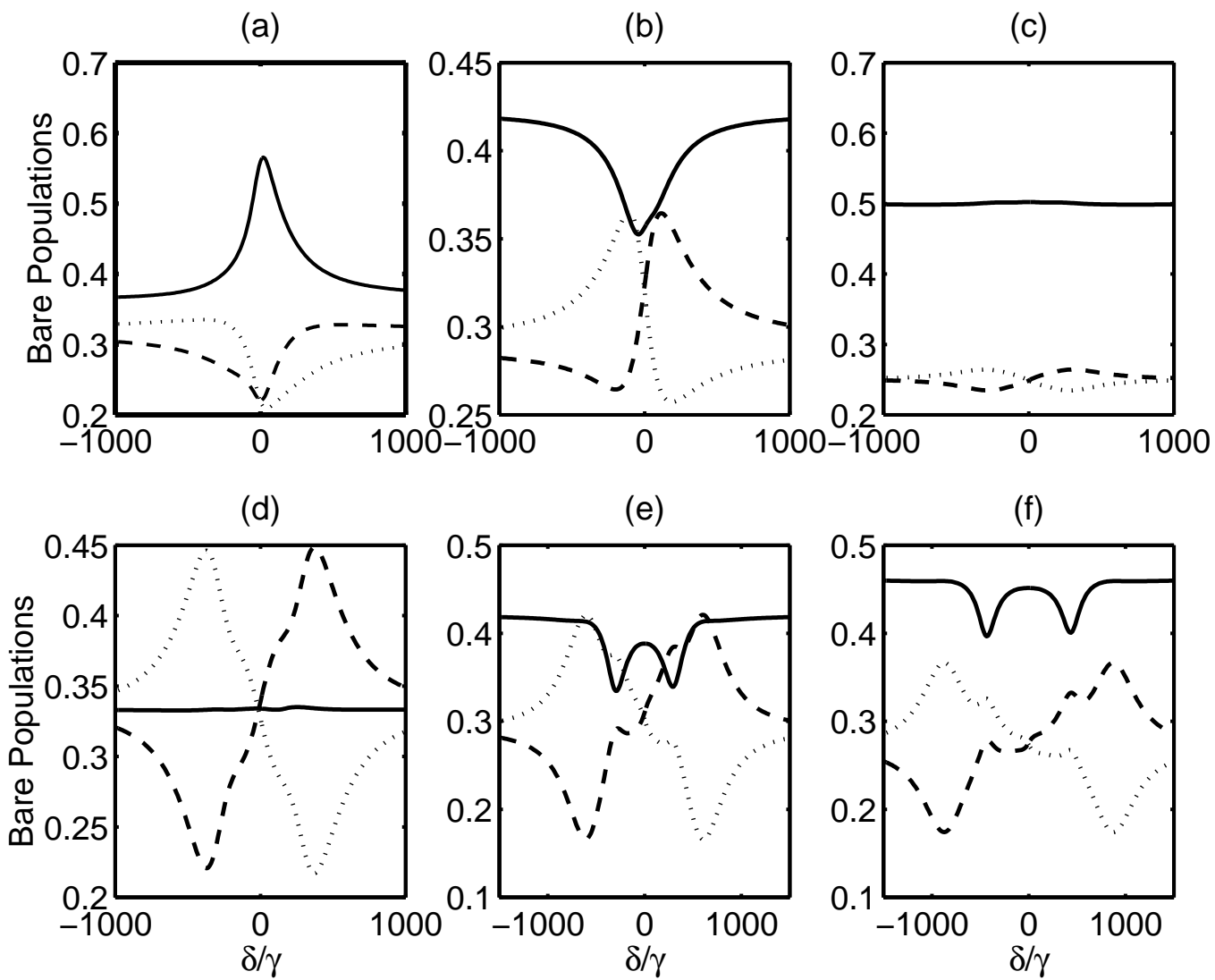
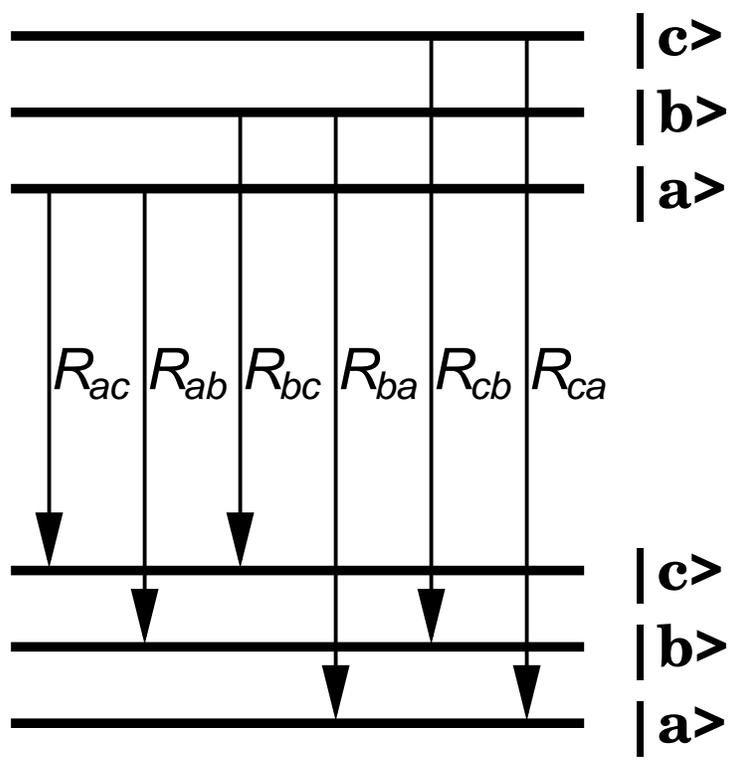
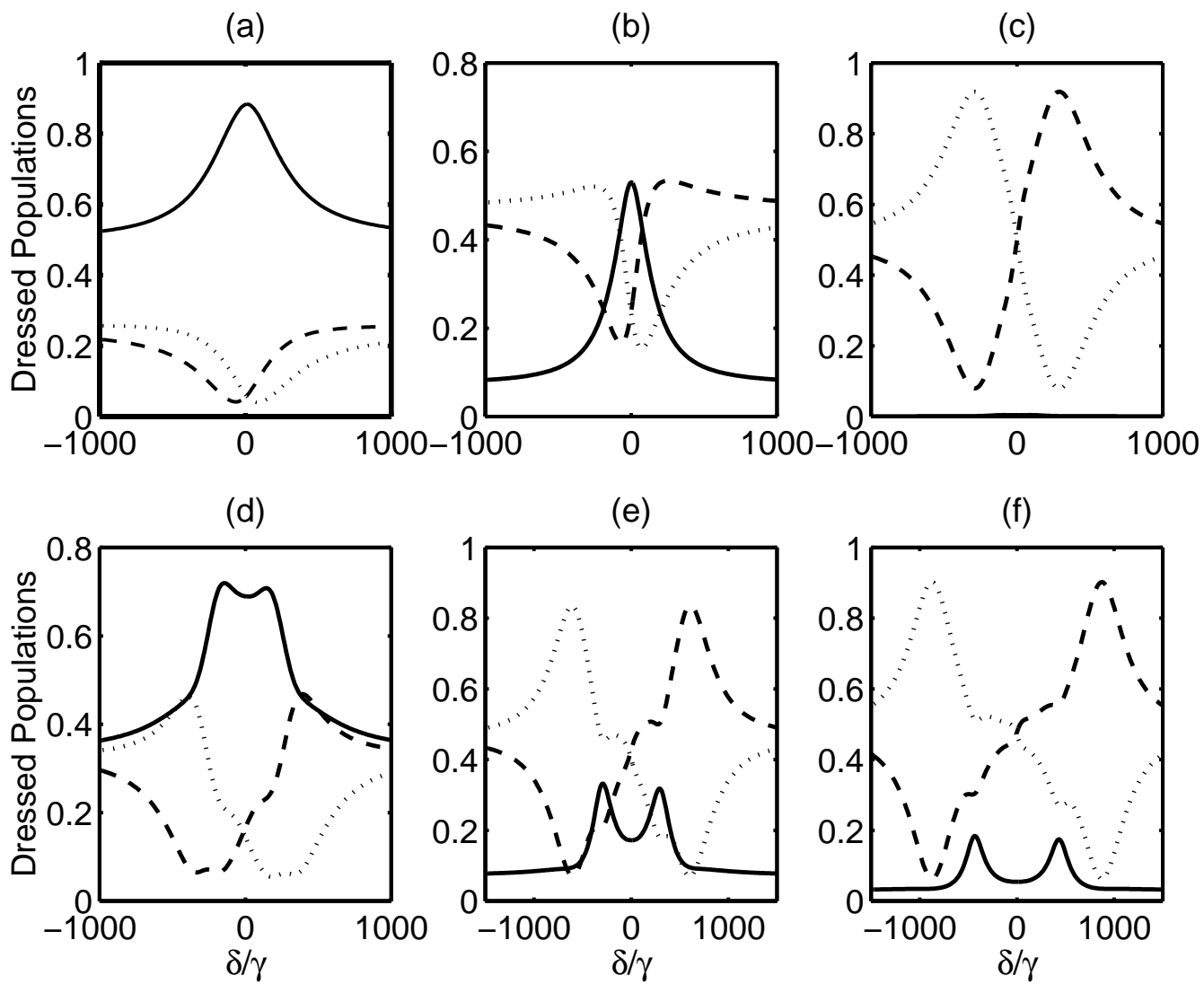
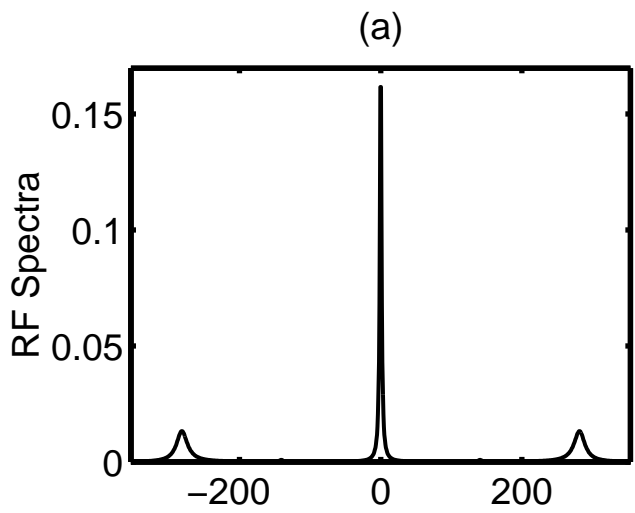
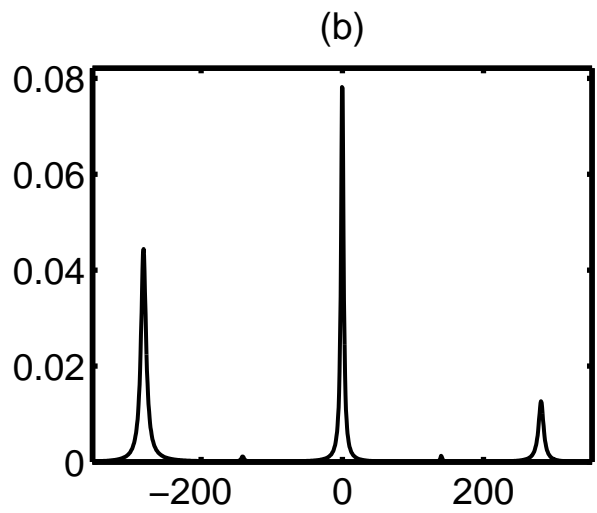
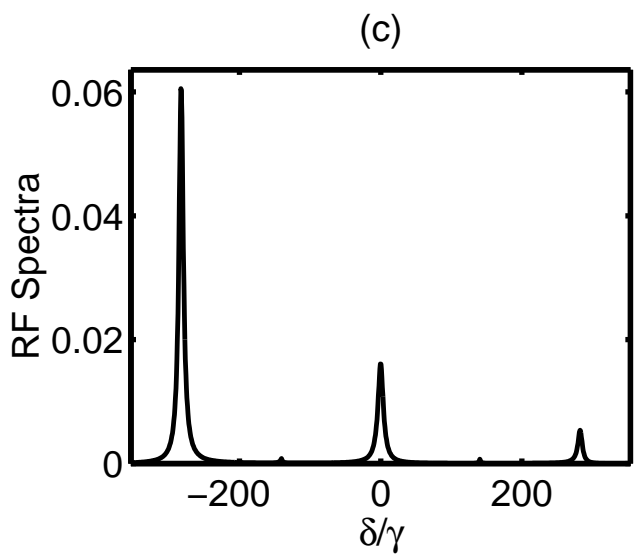
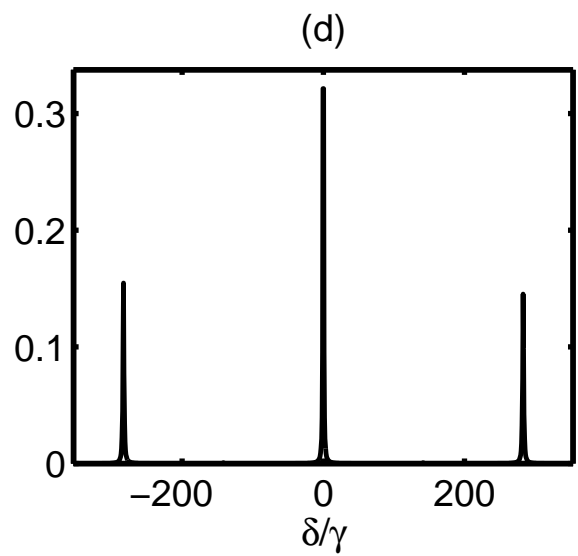
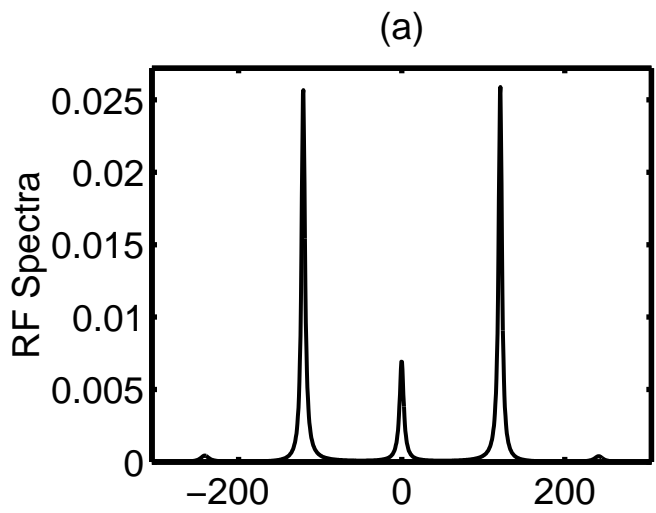
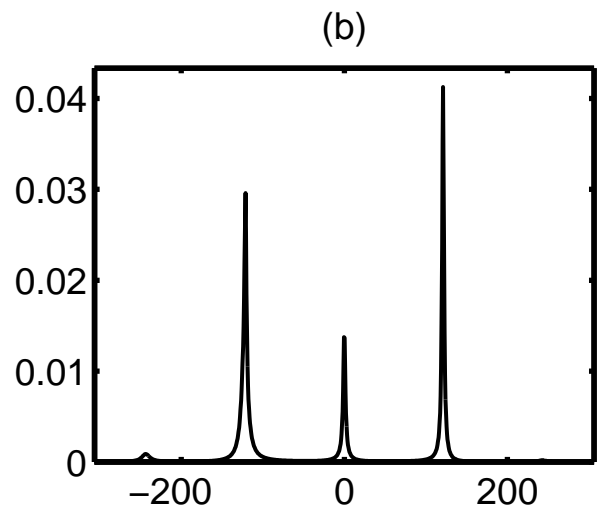
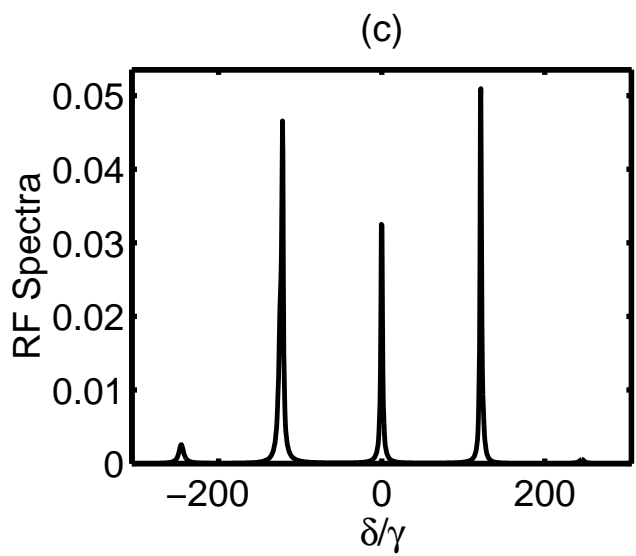
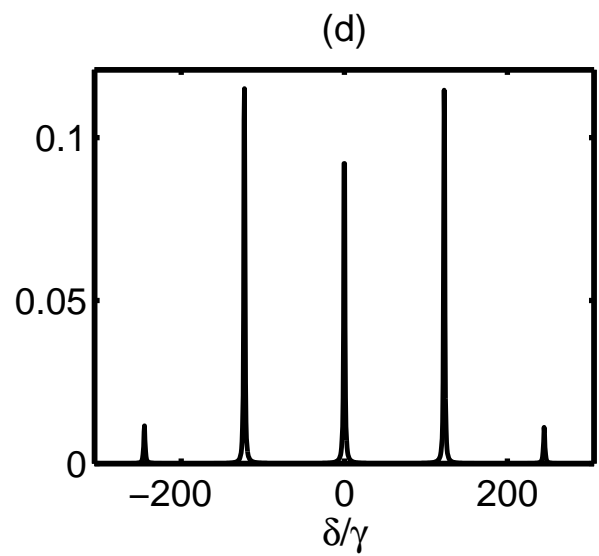
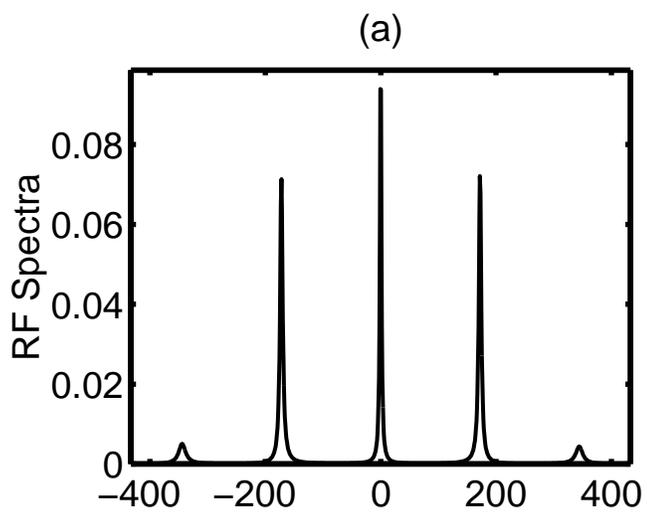
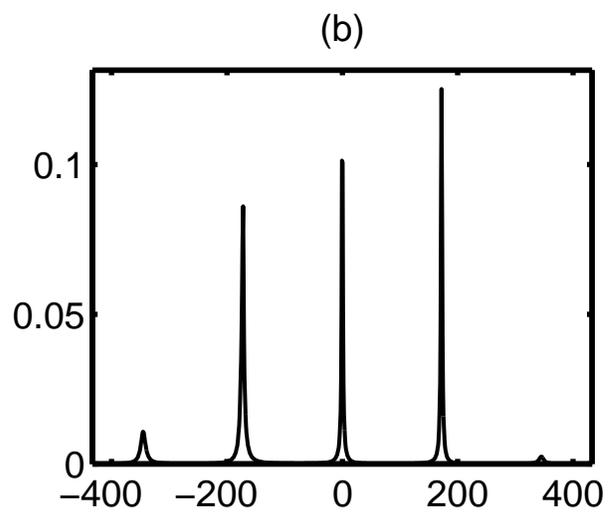
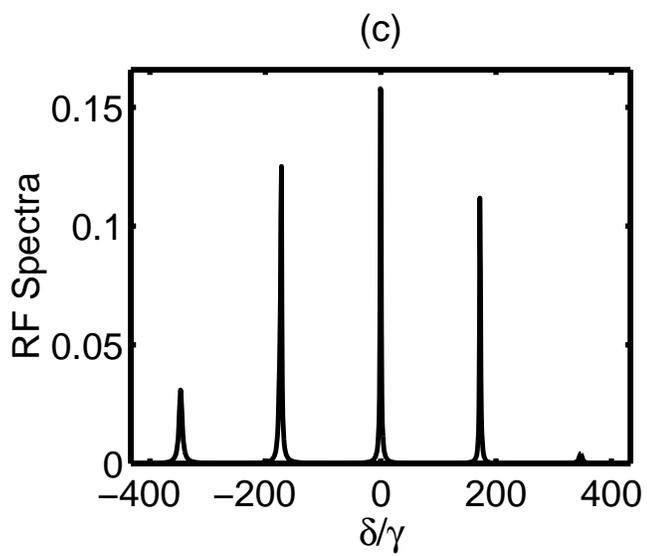
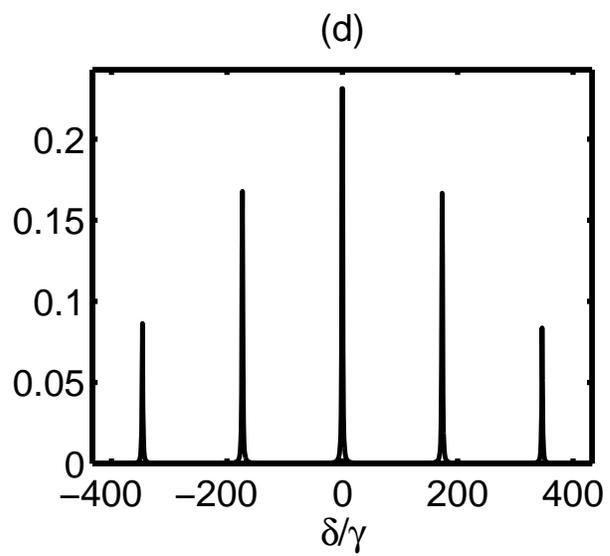
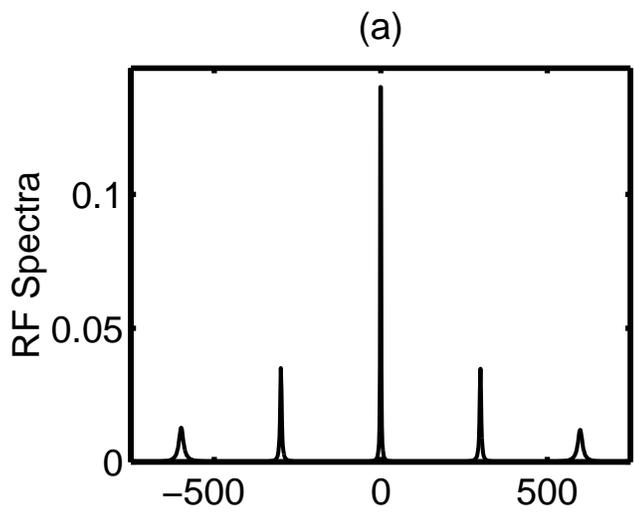
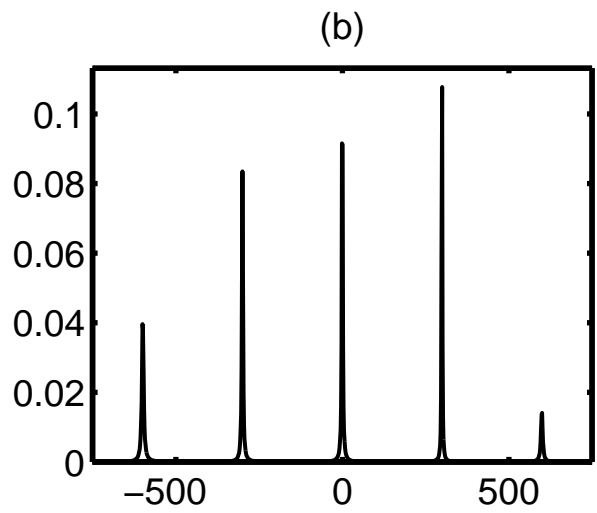
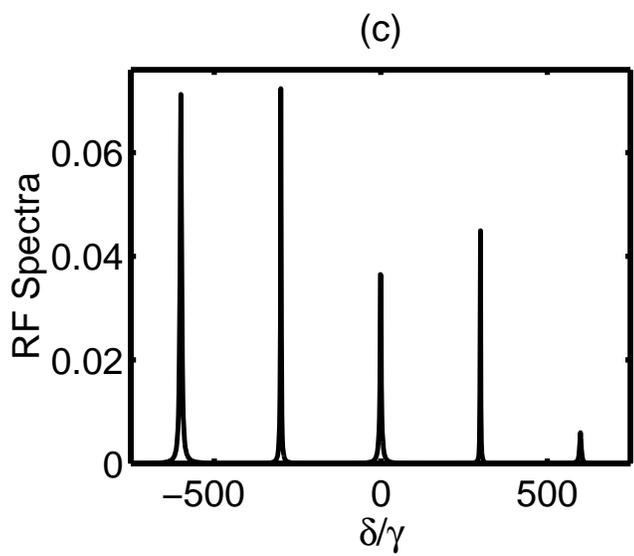
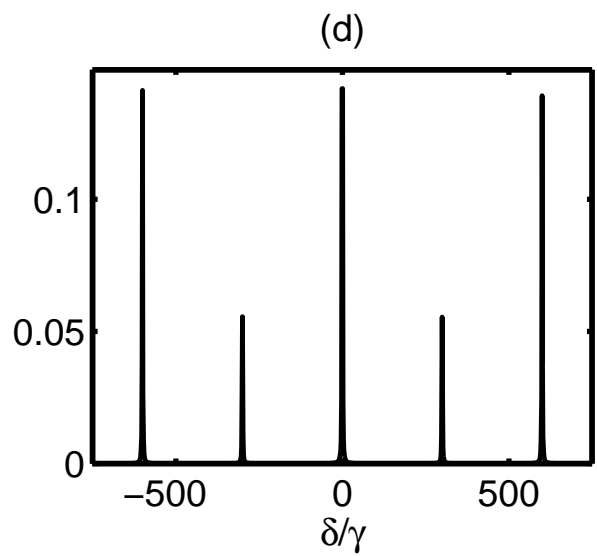
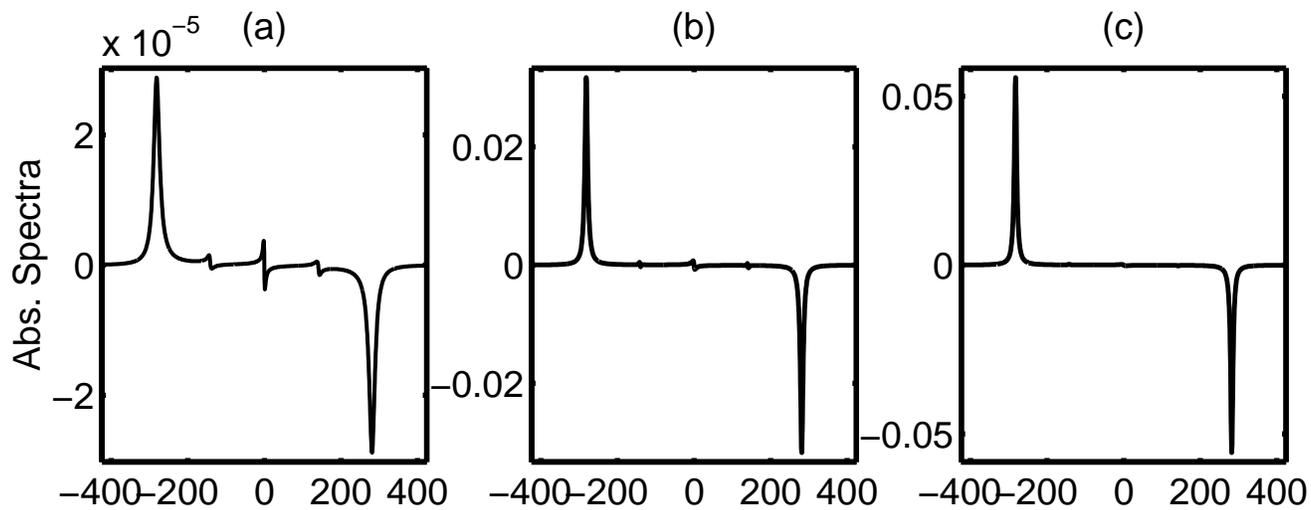
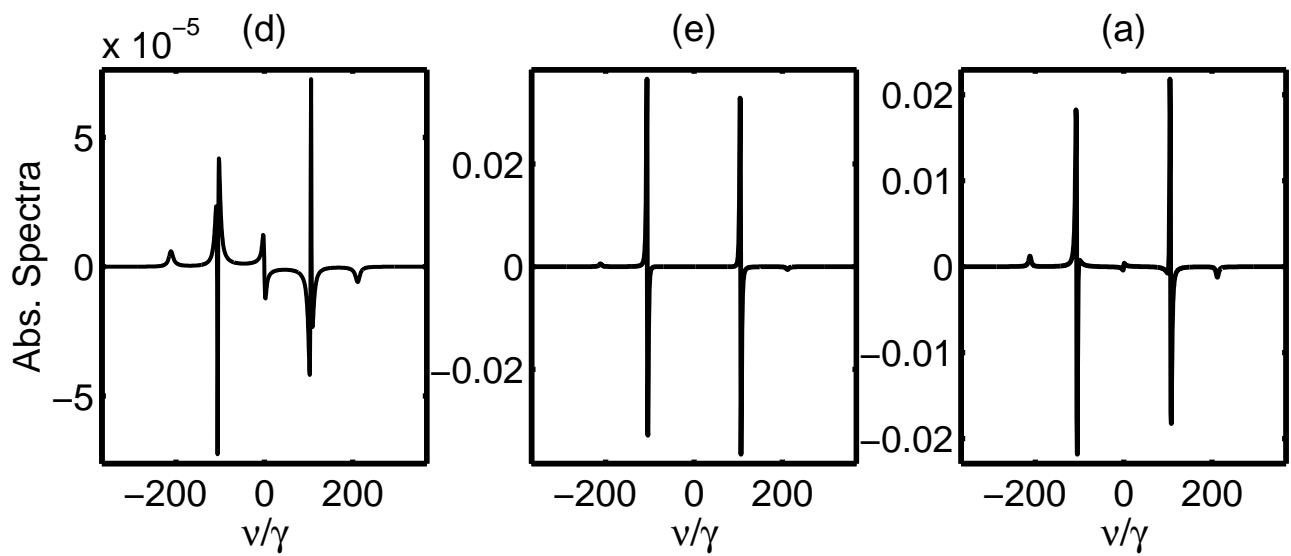
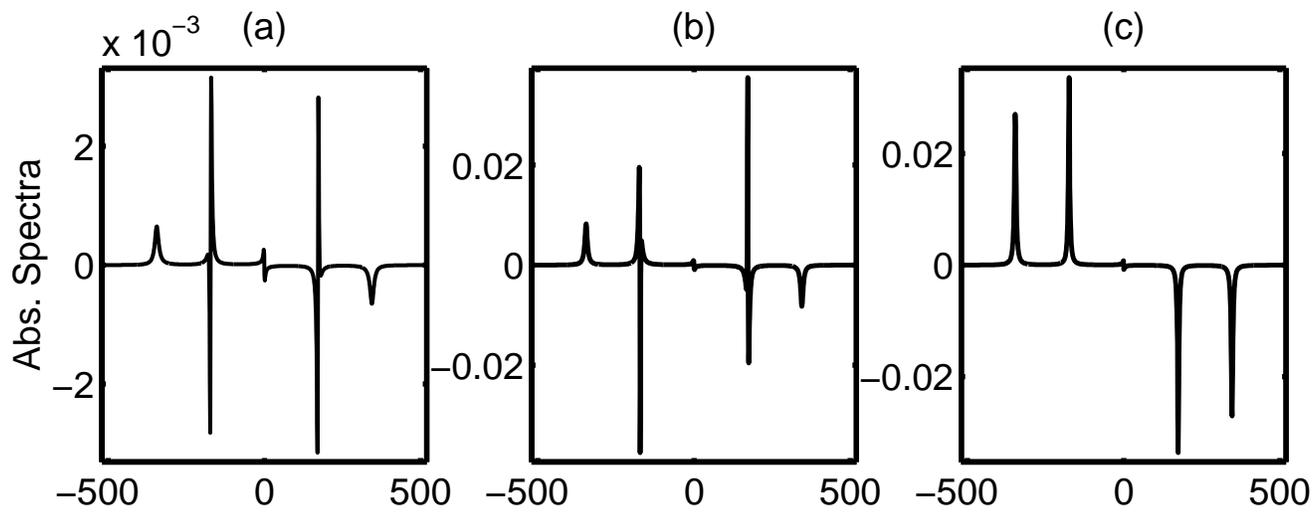
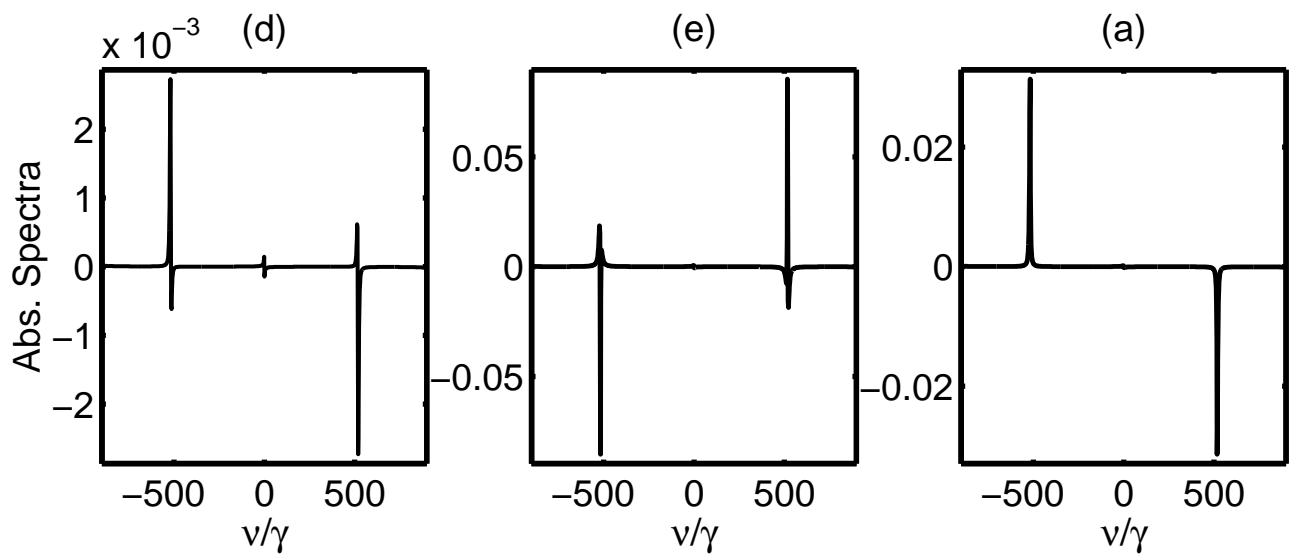
\end{document}